\documentclass[prc,twocolumn,twoside,showpacs,floatfix]{revtex4}

\usepackage{graphicx,color,rotating}
\usepackage{amsmath,amssymb,bm}
\usepackage{dcolumn}

\usepackage{times,txfonts}
\usepackage{pifont}

\newcommand{\eq}[1]{\begin{equation}#1\end{equation}}
\newcommand{\eqmulti}[1]{\begin{equation}\begin{split}#1\end{split}\end{equation}}

\newcommand{\bra}[1]{\ensuremath{\langle{#1}|\,}}
\newcommand{\braas}[1]{\ensuremath{{}_a\langle{#1}|\,}}
\newcommand{\ket}[1]{\ensuremath{\,|{#1}\rangle}}
\newcommand{\ketas}[1]{\ensuremath{\,|{#1}\rangle_a}}
\newcommand{\braket}[2]{\ensuremath{\langle{#1}|{#2}\rangle}}
\newcommand{\braketas}[2]{\ensuremath{{}_a\langle{#1}|{#2}\rangle_a}}

\newcommand{\ketbraas}[2]{\ensuremath{\,|{#1}\rangle_{a\,a}\langle{#2}|\,}}
\newcommand{\matrixe}[3]{\ensuremath{\langle{#1}|\,{#2}\,|{#3}\rangle}}
\newcommand{\matrixeas}[3]{\ensuremath{{}_a\langle{#1}|\,{#2}\,|{#3}\rangle}_a}

\newcommand{\comm}[2]{\ensuremath{[{#1},{#2}]}}

\newcommand{\op}[1]{\ensuremath{#1}}
\newcommand{\adj}[1]{\ensuremath{{{#1}}^{\dag}}}

\newcommand{\clebsch}[6]{ \ensuremath{\left(\!\!
\begin{array}{cc}
 {#1} & \!\!\!\!{#2} \\
 {#4} & \!\!\!\!{#5} 
\end{array}
 \!\!\right|
\left.\!\!\!
\begin{array}{c}
 {#3}\\
 {#6}
\end{array}\!\! \right)}
}

\newcommand{\sixjsym}[6]{ \ensuremath{
\begin{Bmatrix}
 {#1} & \!\!\!{#2} & \!\!\!{#3}\\
 {#4} & \!\!\!{#5} & \!\!\!{#6}
\end{Bmatrix}
}
}
\newcommand{\ninejsym}[9]{
\ensuremath{
\begin{Bmatrix}
 {#1} & \!\!\!{#2} & \!\!\!{#3}\\
 {#4} & \!\!\!{#5} & \!\!\!{#6}\\
 {#7} & \!\!\!{#8} & \!\!\!{#9}
\end{Bmatrix}
}
}

\newcommand{\HOB}[6]{
\ensuremath{
\langle\langle{#1},{#2};{#3}|{#4},{#5}\rangle\rangle_{#6}
}
}

\newcommand{\HO}{\ensuremath{\op{H}}}

\newcommand{\TO}{\ensuremath{\op{T}}}

\newcommand{\pV}{\ensuremath{\vec{p}}}

\newcommand{\piV}{\ensuremath{\vec{\pi}}}

\newcommand{\half}{\ensuremath{\tfrac{1}{2}}}

\newcommand{\elem}[2]{\ensuremath{{}^{#2}\text{#1}}}

\newcommand{\symboltriangle}[1][black]{{\color{#1}\ding{115}}}
\newcommand{\symbolbox}[1][black]{{\color{#1}$\blacksquare$}}
\newcommand{\symboldiamond}[1][black]{{\color{#1}\ding{117}}}
\newcommand{\symbolcircle}[1][black]{{\color{#1}\ding{108}}}
\newcommand{\symbolcross}[1][black]{{\color{#1}\ding{58}}}

\definecolor{FGViolet}{rgb}{0.61,0.32,0.61}
\definecolor{FGDarkBlue}{rgb}{0,0,0.6}
\definecolor{FGBlue}{rgb}{0,0,0.8}
\definecolor{FGLightBlue}{rgb}{0.2, 0.6, 0.8}
\definecolor{FGGreen}{rgb}{0.2,0.7,0.2}
\definecolor{FGLightGreen}{rgb}{0.4,1,0.4}
\definecolor{FGYellow}{rgb}{1,0.95,0}
\definecolor{FGOrange}{rgb}{0.95,0.5,0.1}
\definecolor{FGRed}{rgb}{0.8,0,0}
\definecolor{FGWhite}{rgb}{1,1,1}
\definecolor{FGLightGray}{rgb}{0.8,0.8,0.8}
\definecolor{FGGray}{rgb}{0.5,0.5,0.5}
\definecolor{FGDarkGray}{rgb}{0.3,0.3,0.3}
\definecolor{FGBlack}{rgb}{0,0,0}

\newcommand{\EMAX}{ \ensuremath{E_{3\text{max}}}}

\newcommand{\NNLO}{N${}^2$LO}
\newcommand{\NNNLO}{N${}^3$LO}

\begin{document}

\title{Evolved Chiral NN+3N Hamiltonians for Ab Initio Nuclear Structure Calculations}

\author{Robert Roth}
\email{robert.roth@physik.tu-darmstadt.de}
\author{Angelo Calci}
\email{angelo.calci@physik.tu-darmstadt.de}
\author{Joachim Langhammer}
\email{joachim.langhammer@physik.tu-darmstadt.de}
\author{Sven Binder}
\email{sven.binder@physik.tu-darmstadt.de}

\affiliation{Institut f\"ur Kernphysik, Technische Universit\"at Darmstadt,
64289 Darmstadt, Germany}

\date{\today}

\begin{abstract}  
We discuss the building blocks for a consistent inclusion of chiral three-nucleon (3N) interactions into ab initio nuclear structure calculations beyond the lower p-shell. We highlight important technical developments, such as the similarity renormalization group (SRG) evolution in the 3N sector, a $JT$-coupled storage scheme for 3N matrix elements with efficient on-the-fly decoupling, and the importance truncated no-core shell model with 3N interactions. Together, these developments make converged ab initio calculations with explicit 3N interactions possible also beyond the lower p-shell. We analyze in detail the impact of various truncations of the SRG-evolved Hamiltonian, in particular the truncation of the harmonic-oscillator model space used for solving the SRG flow equations and the omission of the induced beyond-3N contributions of the evolved Hamiltonian. Both truncations lead to sizable effects in the upper p-shell and beyond and we present options to remedy these truncation effects. The analysis of the different truncations is a first step towards a systematic uncertainty quantification of all stages of the calculation.

\end{abstract}


\maketitle

\section{Introduction}

\vspace*{-0.5ex}
Ab initio nuclear structure theory has undergone an amazing development over the past few years, strengthening its role for our understanding of nuclear structure properties on the basis of the strong interaction physics. One of the most active frontiers is the extension of ab initio theories towards heavier nuclei, i.e., beyond the limit around mid p-shell that was characteristic for ab initio approaches a decade ago \cite{BaMi03,WiPi02,PiVa02,NaVa00,NaVa00b}. On the one hand, existing many-body frameworks, such as the no-core shell model (NCSM) \cite{BaNa13,NaQu09,NaGu07} or quantum Monte Carlo methods \cite{GaCa11,GeTe13,PiWi04}, have been improved and extended towards heavier systems. A specific example is the importance truncated NCSM (IT-NCSM) \cite{Roth09,RoNa07}, which extends the domain of NCSM-type calculations into the lower sd-shell. On the other hand, a new generation of many-body methods have been introduced to ab initio nuclear theory, such as coupled-cluster theory \cite{HaPa10,HaPa07,WlDe05,KoDe04}, self-consistent Green's function methods \cite{CiBa13,SoBa13,SoDu11}, or the in-medium similarity renormalization group \cite{HeBi13,HeBo13,TsBo12,TsBo11}, aiming directly at medium-mass nuclei. In many of the recent applications two-nucleon (NN) and three-nucleon (3N) interactions from chiral effective field theory (EFT) are being used as starting point and connection to the underlying physics of the strong interaction \cite{MaEn11,EpHa09}. In comparison to the more phenomenological realistic Hamiltonians used a decade ago, chiral EFT offers a consistent and systematically improvable approach to two-, three-, and multi-nucleon interactions as well as the corresponding electromagnetic and weak operators. From the point of view of nuclear structure observables in light nuclei, already the present generation of chiral NN+3N interaction provides a quantitative description comparable to the best previous realisitc Hamiltonians \cite{WiSt95,Mach01}.

When pushing the ab initio frontier to nuclei beyond the lower p-shell, a particular challenge is the proper inclusion of the 3N interaction at all stages of the calculation. Part of this challenge is the computation and handling of the 3N matrix elements entering the many-body calculations for large model spaces. The huge number of $m$-scheme 3N matrix elements that need to be stored in memory limited the range of previous NCSM calculations \cite{MaAk13,JuMa13,JuNa11,JuNa09}. New developments regarding the computation and handling of 3N matrix elements are mandatory to extend the ab initio frontier beyond the lower p-shell. Similarly, the unitary transformations that are used to enhance the convergence behavior of the many-body calculations have to be extended to the 3N sector. In many of the recent ab initio applications the similarity renormalization group (SRG) is used, since its formal extension to 3N and multi-nucleon interactions is straightforward \cite{JuNa09,RoLa11}. However, the various truncations, e.g., regarding the model spaces used for the numerical solution of the SRG flow equations or the particle rank of the induced many-body contributions, need to be validated. The uncertainties associated with these truncations are expected to become more significant with increasing  particle number. Finally, the many-body approach has to be extended to efficiently include the 3N contributions. In the case of the NCSM this step is straightforward, for methods like coupled-cluster theory it requires a non-trivial extension of the formalism \cite{BiPi13,BiLa13,HaPa07}. Alternatively, one can resort to controlled approximations, such as the normal-ordering approximation discussed in Refs.~\cite{RoBi12,HaPa07}, to partially include 3N interactions while avoiding extensions of the formalism beyond the level of two-body interactions.   

In this technical paper we discuss a chain of key developments enabling the consistent inclusion of chiral 3N interactions into ab initio calculations beyond the lower p-shell, by addressing each of the challenges mentioned above. In Sec.~\ref{sec:me} we discuss the computation of 3N matrix elements starting from a harmonic-oscillator (HO) basis formulated in three-body Jacobi coordinates. We discuss the transformation of the 3N matrix elements to the $JT$-coupled scheme first introduced in Ref.~\cite{RoLa11}, which is used as input for the many-body calculation in conjunction with an efficient on-the-fly decoupling to the $m$-scheme. In Sec.~\ref{sec:srg} we discuss the consistent SRG evolution of the Hamiltonian at the three-body level. We focus on the evolution in a HO representation and introduce new tools, such  as the frequency conversion, to overcome limitations of the HO model space. In Sec.~\ref{sec:itncsm} we discuss the IT-NCSM with explicit 3N interactions and discuss threshold extrapolations of energies and spectroscopic observables. 

Utilizing these tools, in Sec.~\ref{sec:prop}, we critically assess the role of various truncations introduced in the SRG-transformed Hamiltonian. We show ways to remedy truncation errors resulting from the SRG model space and analyze the emergence and the origin of induced beyond-3N interactions. We show that reducing the initial chiral cutoff of the 3N interaction quickly suppresses the SRG-induced beyond-3N contributions leading to an SRG-evolved Hamiltonian with acceptable truncation uncertainties that was already adopted in several applications to medium-mass nuclei \cite{BiPi13,BiLa13,RoBi12,HeBi13,HeBo13,SoBa13,CiBa13}. Finally, in Sec.~\ref{sec:comp} we compare our results to a recent NCSM study \cite{JuMa13} using a more conventional toolchain and discuss different model-space extrapolations.

\section{Three-Body Matrix Elements}
\label{sec:me}

\subsection{Generalities}

The basic input for any many-body approach using a basis expansion within a truncated many-body Hilbert space are appropriate matrix elements of the Hamiltonian. In the context of the NCSM, the underlying basis is given by the eigenstates of the spherical harmonic oscillator (HO), either in the form of $A$-body Slater-determinants of single-particle HO states, the so-called $m$-scheme, or in the form of relative HO states with respect to $A$-body Jacobi coordinates. We will focus on the $m$-scheme formulation, since it is much more convenient when going beyond the lightest nuclei~\cite{BaNa13,NaQu09}. Furthermore, it is more universal and directly applies to other many-body schemes, such as Hartree-Fock calculations, general configuration interaction approaches, or the coupled-cluster method.  

For an $m$-scheme calculation a Hamiltonian containing NN and 3N interactions enters in terms of two- and three-body matrix elements with respect to Slater determinants of two and three HO single-particle states. A prerequisite for a many-body calculation is that these matrix elements can be computed and stored efficiently for sufficiently large basis sizes. 

The computation of these $m$-scheme matrix elements typically involves a multi-step process, which is well established for the two-body matrix elements of the NN interaction. The starting point is an initial representation of the interaction. Typically, one starts with either an operator representation of the interaction or, more conveniently, with a basis representation in a partial-wave decomposed relative-momentum basis $\ket{q (LS) J M; T M_T}_a$, where $q$ is the relative momentum  of the nucleon pair and $\{ (LS) J M; T M_T \}$ are the standard $LS$-coupled partial-wave quantum numbers including total isospin $T$ and isospin projection $M_T$. This basis representation approach has been established as a standard for the chiral NN interactions~\cite{EnMa03,EpGl05,EkBa13}. In a first step, we compute relative HO matrix elements for the basis $\ket{N (LS) J M; T M_T}_a$ with radial HO quantum number $N$ using a simple basis transformation. In a 
second step, the relative HO matrix elements can be converted through a Talmi-Moshinsky transformation plus angular momentum recouplings \cite{KaKa01,RoPa06} into $m$-scheme matrix elements with respect to the antisymmetrized two-body states $\ket{n_al_aj_am_am_{ta}; n_bl_bj_bm_bm_{tb}}_a$ with single-particle HO quantum numbers. In order to reduce the storage requirements for the  two-body matrix elements and to exploit the symmetries of the two-body interaction, one generally does not store $m$-scheme matrix elements directly, but a simple $JT$-coupled form with respect to the basis states $\ket{ n_a l_a ; n_b l_b ; (j_a j_b) J M ; (\half \half) T M_T }_a$. The decoupling to pure $m$-scheme matrix elements is done on the fly during the many-body calculation. 

For the 3N interaction, we follow the exactly same route, though each of the steps is significantly more involved. Again, the 3N interaction is initially given in an operator form or in a partial-wave decomposed Jacobi-momentum basis. In a first step, the latter can be transformed into a partial-wave Jacobi-coordinate HO basis, which also gives an easy handle on antisymmetrization. Then in a second step, we could transform from Jacobi to $m$-scheme HO matrix elements through a sequence of two Talmi-Moshinsky transformations and recouplings. This strategy was used in previous large-scale applications of chiral 3N interaction in the NCSM, see e.g. Ref.~\cite{NaGu07,JuNa09,MaVa11}. We propose to use $JT$-coupled three-body matrix elements for a more efficient storage and retrieval combined with an on-the-fly decoupling during the many-body calculation~\cite{RoLa11}, in complete analogy to the standard procedure for two-body matrix elements. We will discuss the details and the advantages of this scheme in the following.

\subsection{Initial 3N matrix elements}
\label{sec:3nme_initial}

For the chiral 3N interaction, the computation of initial partial-wave decomposed relative matrix elements can be challenging already. To be specific, we consider three-body matrix elements with respect to the two Jacobi momenta $\piV_1$ and $\piV_2$ in the three-body system, defined by~\cite{NaKa00,Navr07}
\eq{
\piV_{1} = \tfrac{1}{\sqrt{2}} (\pV_a - \pV_b)
\;,\quad
\piV_{2}  = \sqrt{\tfrac{2}{3}} \Big[\tfrac{1}{2}(\pV_a + \pV_b) - \pV_c \Big]
}
where $\pV_{a,b,c}$ are the single-particle momenta of the three nucleons. The Jacobi momentum $\piV_0$ characterizing the center-of-mass motion is irrelevant for the description of the intrinsic dynamics. We systematically use numeric indices for quantities defined with respect to relative Jacobi coordinates and latin indices for quantities defined with respect to single-particle coordinates. For example, $L_1$ denotes a relative orbital angular-momentum quantum number with respect to the first Jacobi coordinate $\piV_1$, whereas $l_a$ denotes a single-particle orbital angular momentum. As a general rule, we use capital letters for angular momentum, spin and isospin quantum numbers that involve more than one particle and lower-case letters for single-particle quantum numbers.

The starting point for the following calculation is a partial-wave representation of the Jacobi-momentum basis in the three-nucleon system. Using a $J_1J_2$-coupling scheme for the two total angular momenta $J_1$ and $J_2$ associated with the Jacobi momenta $\pi_1$ and $\pi_2$ we write the basis states as
\eq{\label{eq:jme3_mom}
  \ket{\pi_1 \pi_2;\alpha} 
  = \ket{\pi_1 \pi_2; [(L_1 S_1) J_1, (L_2\half) J_2] J_{12}; (T_1 \half) T_{12} } 
}
with $\alpha = \{ [(L_1 S_1) J_1, (L_2\half) J_2] J_{12}; (T_1 \half) T_{12} \}$ as a collective index for all angular momentum, spin and isospin quantum numbers defining the partial wave. We omit the projection quantum numbers $M_{12}$ and $M_{T12}$ for brevity. Note that these basis states have a well-defined transposition symmetry only with respect to the particles $a$ and $b$, we will discuss the complete antisymmetrization in the context of the Jacobi-HO matrix elements in Sec.~\ref{sec:3nme_jacobiho}.  

The computation of matrix elements of the chiral 3N interaction in this basis is the first step. For 3N interactions at \NNLO\ there are only five different momentum-spin-isospin structures, for which a partial wave decomposition can be performed explicitly. This is discussed in detail in Refs.~\cite{HuWi97,EpNo02} and in Ref.~\cite{Navr07} for different formulations of the regulators.

For the chiral 3N interaction at \NNNLO\ the situation changes radically. Recently, the derivation of cartesian momentum-space structures of the 3N interaction at \NNNLO\ was completed~\cite{BeEp08,BeEp11}. In view of the many different momentum-spin-isospin operators involved, a manual partial-wave decomposition is hardly feasible. Therefore, an automatized partial-wave decomposition was recently proposed by Skibi\'nski et al.~\cite{SkGo11}, which uses numerical integrations over five angular variables to extract partial-wave Jacobi-momentum matrix elements. As a result tabulated numerical values of the matrix elements on a four-dimensional grid of Jacobi momenta will be available for subsequent calculations. The partial-wave decomposition is  computationally quite expensive and there is an ongoing collaborative effort within the LENPIC \footnote{Low-Energy Nuclear Physics International Collaboration (LENPIC), see http://www.lenpic.org} collaboration to generate those matrix elements for the chiral interaction at \NNNLO\ for use in nuclear structure calculations.

\subsection{Jacobi-HO matrix elements}
\label{sec:3nme_jacobiho}

When aiming at many-body calculations using an HO basis, it is convenient to transform the three-body Jacobi matrix elements into an HO representation right away. We use a partial-wave Jacobi-HO basis of the form
\eq{ \label{eq:jacobiHO_partial_antisym}
  \ket{N_1 N_2;\alpha} 
  = \ket{N_1 N_2; [(L_1 S_1) J_1, (L_2\half) J_2] J_{12}; (T_1 \half) T_{12} } 
}
with radial HO quantum numbers $N_1$ and $N_2$ defined with respect to the first and second Jacobi coordinate and the collective partial-wave index $\alpha$ as in the Jacobi-momentum representation. The transformation of three-body matrix elements from the $\ket{\pi_1 \pi_2;\alpha}$ to the $\ket{N_1 N_2;\alpha}$ basis is straight forward.

Within the Jacobi-HO representation we can also perform the complete antisymmetrization of the three-body matrix elements in a convenient manner. Following Refs.~\cite{NoNa06,Navr07} we denote antisymmetrized Jacobi-HO states as $\ket{E_{12} i J^{\pi}_{12} T_{12}}_a$, where $E_{12} = (2 N_1 + L_1) + (2 N_2 + L_2)$ is the principal HO quantum number of the Jacobi-HO state, $J^{\pi}_{12}$ is the total angular momentum and parity of the relative motion and $T_{12}$ the total isospin. These are the only good quantum numbers of the antisymmetrized Jacobi-HO basis. The index $i$ labels the different antisymmetrized basis states that emerge for given $E_{12}$, $J^{\pi}_{12}$, and $T_{12}$---it does not correspond to a physically meaningful quantum number. The transformation to the antisymmetrized Jacobi-HO basis can be written as 
\eqmulti{ \label{eq:jacobiHO_antisym}
  \ketas{E_{12} i J^{\pi}_{12} T_{12}}
  = \sum_{N'_1,N'_2,\alpha'}
     &\delta_{E_{12},(2 N'_1 + L'_1) + (2 N'_2 + L'_2)} 
     \delta_{J^{\pi}_{12}, J'^{\pi'}_{12}}
     \delta_{T_{12},T'_{12}} \\ 
     \times\; & C^{i}_{N'_1 N'_2 \alpha'} 
     \ket{N'_1 N'_2; \alpha' } 
}
where the overlap of the non-antisymmetrized and the antisymmetrized Jacobi-HO states defines so-called coefficients of fractional parentage (CFP)~\cite{NaOr03,NoNa06,Navr07}
\eq{ \label{eq:CFP}
   C^{i}_{N_1 N_2 \alpha} 
   = \braket{N_1 N_2; \alpha}{E_{12} i J^{\pi}_{12} T_{12}}_a
}
with $E_{12} = (2 N_1 + L_1) + (2 N_2 + L_2)$. The numerical values of the CFPs can be determined by solving the eigenvalue problem of the antisymmetrization operator $\mathcal{A}$ in the Jacobi-HO basis $\ket{N_1 N_2; \alpha }$. This matrix exhibits a block structure in $E_{12}$, $J^{\pi}_{12}$, and $T_{12}$, indicating that these are good quantum numbers in both representations. The eigenvectors of the matrix in each $(E_{12}, J^{\pi}_{12}, T_{12})$ block that belong to the degenerate subspace to the eigenvalue $1$ define the CFPs with $i$ as a degeneracy index~\cite{NaKa00}. The Kronecker deltas in Eq.~\eqref{eq:jacobiHO_antisym} reduce the summations to the $(E_{12}, J^{\pi}_{12}, T_{12})$ block defined through the left-hand side.

Transformation \eqref{eq:jacobiHO_antisym} is a highly efficient way to project the Jacobi-HO states $\ket{N_1 N_2;\alpha}$ onto a complete orthonormalized basis of antisymmetric states. The numerical simplicity of the transformation to the antisymmetrized basis is the main advantage of working with a Jacobi-HO basis as compared to the Jacobi-momentum representation \cite{Hebe12}.

\subsection{Transformation to $JT$-coupled matrix elements}

The most demanding step in the preparation of three-body matrix elements for many-body calculations is their transformation from the Jacobi-HO basis into a three-body Slater-determinant basis of HO single-particle states, also called $m$-scheme states. We are interested in matrix elements with respect to an antisymmetrized $JT$-coupled three-body basis composed of HO single-particle states
\eqmulti{ \label{eq:JT_coupled_states}
  &\ketas{ \tilde{a} \tilde{b} \tilde{c}; J_{ab} J; T_{ab} T } = \\
  &\quad = \ketas{n_a l_a n_b l_b n_c l_c ; [(j_a j_b) J_{ab}, j_c] J ; [(\half \half) T_{ab}, \half] T}
}
where $\tilde{a}=\{n_a,l_a,j_a\}$, etc. is a short hand for the radial and angular momentum single-particle quantum numbers and the projection quantum numbers $M$ and $M_T$ are omitted. These antisymmetrized states can be generated from $JT$-coupled product states by applying the antisymmetization operator $\mathcal{A}$ explicitly
\eq{
  \ketas{ \tilde{a} \tilde{b} \tilde{c}; J_{ab} J; T_{ab} T } = 
  \sqrt{6}\; \mathcal{A}\; \ket{ \tilde{a} \tilde{b} \tilde{c}; J_{ab} J; T_{ab} T } \; ,
}
where we introduce a normalization factor and, thus, define $\mathcal{A}$ as projection operator. To connect the non-antisymmetrized $JT$-coupled basis with the center-of-mass frame relative Jacobi-HO states, we have to augment the latter with an explicit center-of-mass component $\ket{N_{cm} L_{cm}}$ again using the HO basis. Starting from the non-antisymmetrized Jacobi-HO states \eqref{eq:jacobiHO_partial_antisym} we define
\eq{ \label{eq:JT_coupled_CMframe_states}
 \ket{N_1 N_2 ; \alpha ; N_{cm}L_{cm} ; J} =
 \{ \ket{N_1 N_2; \alpha} \otimes \ket{N_{cm}L_{cm}} \}^J
}
where $J$ results from the coupling of $J_{12}$ for the relative motion with $L_{cm}$ for the center of mass. As before, all projection quantum numbers are suppressed for brevity. The overlap of the $JT$-coupled laboratory-frame states $\ket{ \tilde{a} \tilde{b} \tilde{c}; J_{ab} J; T_{ab} T }$ with the Jacobi states~\eqref{eq:JT_coupled_CMframe_states} defines the transformation coefficient
\eqmulti{ \label{eq:Tcoefficient_definition}
  &T^{\tilde{a} \tilde{b} \tilde{c} J_{ab} J}_{N_1 N_2 \alpha N_{cm} L_{cm}} = 
  T^{n_a l_a j_a n_b l_b j_b n_c l_c j_c\, J_{ab} J}_{N_1 L_1 S_1 J_1\,N_2 L_2 J_2\, J_{12}\, N_{cm} L_{cm}}= \\
  &\quad 
  = \braket{ N_1 N_2 ; \alpha ; N_{cm}L_{cm} ; J }{ \tilde{a}\, \tilde{b}\, \tilde{c}; J_{ab} J; T_{ab} T }\; . 
}
This overlap is independent of the isospin quantum numbers and non-vanishing only for $T_1 = T_{ab}$ and $T_{12} = T$. Through multiple angular-momentum recouplings and Talmi-Moshinsky transformations one can work out the following analytic form of the $T$ coefficients, as discussed in Ref.~\cite{NoNa06}
\eqmulti{ \label{eq:Tcoefficients}
  &T^{n_a l_a j_a n_b l_b j_b n_c l_c j_c J_{ab} J}_{N_1 L_1 S_1 J_1 N_2 L_2 J_2 J_{12} N_{cm} L_{cm}} = \\
  &\quad 
  = \sum_{ \mathcal{N}, \mathcal{L} } \sum_{L_{ab}} \sum_{L} \sum_{S_{12}} \sum_{L_{12}} \sum_{\Lambda} \\
  &\quad\quad\, 
  \delta_{2n_a + l_a + 2n_b + l_b + 2n_c + l_c, 2N_{cm} + L_{cm} + 2 N_1 + L_1 + 2 N_2 + L_2} \\
  &\quad
  \times  
  (-1)^{l_c + \Lambda + L_{ab} + L + S_{12} + L_1 + J} \\
  &\quad
  \times 
  \hat{\jmath}_a \hat{\jmath}_b \hat{\jmath}_c \hat{J}_{ab} \hat{J} \hat{J}_{1} \hat{J}_{2} \hat{S}_1 
  \hat{S}_{12}^2 \hat{L}_{ab}^2 \hat{L}^2 \hat{L}_{12}^2 \hat{\mathcal{L}}^2 \hat{\Lambda}^2  \\
  &\quad
  \times 
  \HOB{\mathcal{N} \mathcal{L}}{N_1 L_1}{L_{ab}}{n_b l_b}{n_a l_a}{1} \\
  &\quad
  \times 
  \HOB{N_{cm} L_{cm}}{N_2 L_2}{\Lambda}{\mathcal{N} \mathcal{L}}{n_c l_c}{2} \\
  &\quad
  \times 
  \ninejsym{l_a}{l_b}{L_{ab}}{\half}{\half}{S_1}{j_a}{j_b}{J_{ab}}
  \ninejsym{L_{ab}}{l_c}{L}{S_1}{\half}{S_{12}}{J_{ab}}{j_c}{J}
  \ninejsym{L_1}{L_2}{L_{12}}{S_1}{S_2}{S_{12}}{J_1}{J_2}{J_{12}} \\
  &\quad
  \times 
  \sixjsym{l_c}{\mathcal{L}}{\Lambda}{L_1}{L}{L_{ab}}
  \sixjsym{L_{cm}}{L_2}{\Lambda}{L_1}{L}{L_{12}}
  \sixjsym{L_{cm}}{L_{12}}{L}{S_{12}}{J}{J_{12}}
}   
with the short-hand $\hat{x}=\sqrt{2x+1}$. Due to the change of the underlying coordinate system for the description of the three nucleons, two harmonic-oscillator brackets (HOBs) $\langle\langle\ldots|\ldots\rangle\rangle_{1,2}$ appear~\cite{KaKa01}. The HOBs always require a coupling of orbital angular momenta, which implies various angular-momentum recouplings, resulting in the $6j$- and $9j$-symbols. The $\mathcal{N}$ summation can be eliminated using the energy-conservation property of the first HOB.

We now have all components to formulate the matrix elements of the three-body operator $V$ in the antisymmetrized $JT$-coupled basis
\eqmulti{ \label{eq:JTcoupled_matrixelement_pre}
  &\matrixeas{ \tilde{a} \tilde{b} \tilde{c}; J_{ab} J; T_{ab} T }{ V }
             { \tilde{a}' \tilde{b}' \tilde{c}'; J'_{ab} J; T'_{ab} T } = \\
  &\quad = 6 \; 
  \matrixe{ \tilde{a} \tilde{b} \tilde{c}; J_{ab} J; T_{ab} T }{ \mathcal{A} V \mathcal{A} }
          { \tilde{a}' \tilde{b}' \tilde{c}'; J'_{ab} J; T'_{ab} T } \;,
}
where we again omit all projection quantum numbers. We can express the antisymmetrization operator using the antisymmetrized Jacobi-HO basis, augmented by a HO center-of-mass part analogously to Eq.~\eqref{eq:JT_coupled_CMframe_states},
\eqmulti{
  \mathcal{A} 
  &= \sum_{E_{12},i,J^{\pi}_{12},T_{12}} \sum_{N_{cm},L_{cm}} \sum_{J} \\
  &\quad \ketbraas{E_{12} i J^{\pi}_{12} T_{12}; N_{cm} L_{cm}; J}{E_{12} i J^{\pi}_{12} T_{12}; N_{cm} L_{cm}; J}\; .
} 
Plugging this into Eq.~\eqref{eq:JTcoupled_matrixelement_pre} and inserting additional resolutions of the unit operator in the non-antisymmetrized Jacobi-HO basis \eqref{eq:jacobiHO_partial_antisym} using
\eqmulti{
  &\braket{N_1 N_2; \alpha; N_{cm} L_{cm}; J}{E'_{12} i J'^{\pi}_{12} T'_{12}; N'_{cm} L'_{cm}; J'}_a 
  = C^{i}_{N_1 N_2 \alpha} \\
  &\quad\times\delta_{(2 N_1 + L_1) + (2 N_2 + L_2), E'_{12}} 
  \delta_{J^{\pi}_{12}, J'^{\pi'}_{12}}
    \delta_{T_{12},T'_{12}} 
    \delta_{N_{cm},N'_{cm}}
    \delta_{L_{cm},L'_{cm}}
    \delta_{J,J'}
}
as well as the definition of the $T$ coefficients~\eqref{eq:Tcoefficient_definition}, we arrive at the final transformation equation
\eqmulti{ \label{eq:JTcoupled_matrixelement}
  &\matrixeas{\tilde{a} \tilde{b} \tilde{c}; J_{ab} J; T_{ab} T }{ V }{ \tilde{a}' \tilde{b}' \tilde{c}'; J'_{ab} J; T'_{ab} T } = \\
  &\quad
  =
  6 \sum_{N_1,N_2,\alpha} \sum_{N'_1,N'_2,\alpha'} \sum_{N_{cm},L_{cm}} \sum_{i,i'} \\
  &\quad\quad\,\delta_{T_{ab},T_1} \delta_{T'_{ab},T'_1} \delta_{T,T_{12}} \delta_{T,T'_{12}} \delta_{J_{12},J'_{12}} \\
  &\quad\times
  T^{\tilde{a}\, \tilde{b}\, \tilde{c}\, J_{ab} J}_{N_1 N_2 \alpha N_{cm} L_{cm}} \;
  T^{\tilde{a}'\, \tilde{b}'\, \tilde{c}'\, J'_{ab} J}_{N'_1 N'_2 \alpha' N_{cm} L_{cm}} \;
  C^{i}_{N_1 N_2 \alpha} \;
  C^{i'}_{N'_1 N'_2 \alpha'} \\[3pt]
  &\quad\times 
  \matrixeas{E_{12} i J^{\pi}_{12} T_{12}}{ V }{ E'_{12} i' J^{\pi}_{12} T_{12}}
}
with $E_{12} = (2 N_1 + L_1) + (2 N_2 + L_2)$ and $E'_{12} = (2 N'_1 + L'_1) + (2 N'_2 + L'_2)$. The first four Kronecker deltas eliminate the isospin summations contained in the $\alpha,\alpha'$ sums and ensure $T_1 = T_{ab}$, $T_{12} = T$, etc. 

The transformation given by Eq.~\eqref{eq:JTcoupled_matrixelement} is computationally demanding, mainly because of the sheer number of relevant $T$ coefficients. Some of the computational aspects and limitations for evaluating this transformation are discussed in Sec.~\ref{subsec:computational_strategy}.

\subsection{Decoupling to $m$-scheme}

For many-body calculations using an $m$-scheme basis, it is crucial to efficiently obtain the three-body matrix elements in a corresponding uncoupled or $m$-scheme representation
\eqmulti{ \label{eq:mscheme_states}
  \ketas{ a b c } 
  = \ketas{n_a l_a j_a m_{ja} m_{ta};  n_b l_b j_b m_{jb} m_{tb}; n_c l_c j_c m_{jc} m_{tc}}\, ,
}
where $a=\{n_a l_a j_a m_{ja} m_{ta}\}$ is a short hand for the single-particle quantum numbers, including all projection quantum numbers.
Thus, the final step in the computational scheme is the complete decoupling of the antisymmetrized $JT$-coupled matrix elements to obtain pure antisymmetrized $m$-scheme matrix elements 
\eqmulti{ \label{eq:Mscheme_matrixelement}
  &\matrixeas{abc}{V}{a'b'c'}
  = \sum_{J_{ab},J'_{ab},J} \sum_{T_{ab},T'_{ab},T} \\
  &\quad\times
  \clebsch{j_a}{j_b}{J_{ab}}{m_a}{m_b}{M_{ab}}
  \clebsch{J_{ab}}{j_c}{J}{M_{ab}}{m_c}{M}
  \clebsch{\half}{\half}{T_{ab}}{m_{ta}}{m_{tb}}{M_{Tab}}
  \clebsch{T_{ab}}{\half}{T}{M_{Tab}}{m_{tc}}{M_T} \\
  &\quad\times
  \clebsch{j'_a}{j'_b}{J'_{ab}}{m'_a}{m'_b}{M'_{ab}}
  \clebsch{J'_{ab}}{j'_c}{J}{M'_{ab}}{m'_c}{M}
  \clebsch{\half}{\half}{T'_{ab}}{m'_{ta}}{m'_{tb}}{M'_{Tab}}
  \clebsch{T'_{ab}}{\half}{T}{M'_{Tab}}{m'_{tc}}{M_T} \\
  &\quad\times
  \matrixeas{\tilde{a} \tilde{b} \tilde{c}; J_{ab} J; T_{ab} T }{ V }{ \tilde{a}' \tilde{b}' \tilde{c}'; J'_{ab} J; T'_{ab} T }
}
with all $M$ and $M_T$ quantum numbers determined by sums of the single-particle $m$ and $m_t$ quantum numbers, e.g., $M_{ab}=m_a+m_b$.
This decoupling is trivial and requires only Clebsch-Gordan coefficients. Therefore, the decoupling can be easily and efficiently done on the fly during the many-body calculation.

\subsection{Computational strategy}
\label{subsec:computational_strategy}

After discussing the formal steps for the calculation of the three-body matrix elements entering NCSM-type many-body calculations, we would like to address a few computational aspects, since they are crucial for practical applications and set the limits for present ab initio calculations. 

\begin{figure}
\includegraphics[width=0.8\columnwidth]{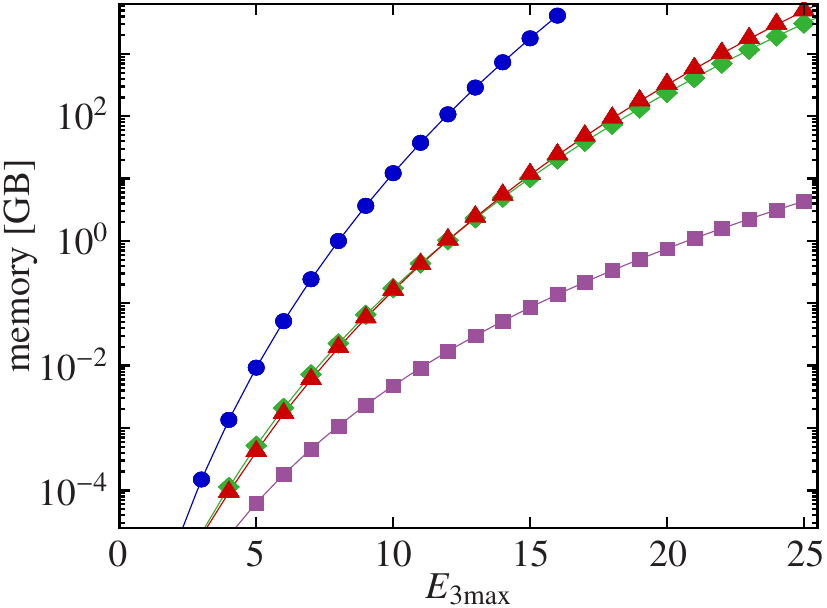}
\caption{(color online)
Memory required to store the $T$-coefficients (\symboldiamond[FGGreen]), as well as the three-body matrix elements in the antisymmetrized-Jacobi (\symbolbox[FGViolet]), $JT$-coupled (\symboltriangle[FGRed]), and $m$-scheme (\symbolcircle[FGBlue]) representation as function of the  maximum three-body energy quantum number $\EMAX$. All quantities are assumed to be single-precision floating point numbers.  
}
\label{fig:memoryneeds}
\end{figure}
 
The calculation of three-body matrix elements is a prime example for the 'recompute versus store' paradigm. In many NCSM applications including chiral 3N interactions~\cite{NaGu07,MaVa11,JuMa13}, the complete set of $m$-scheme matrix elements~\eqref{eq:Mscheme_matrixelement} was computed and stored before the actual many-body calculation. As mentioned earlier, the sheer number of three-body $m$-scheme matrix elements sets a severe limit to the model-space sizes that are accessible with this approach. This is illustrated in Fig.~\ref{fig:memoryneeds} which shows the memory needed to store $m$-scheme matrix elements of the 3N interaction exploiting all basic symmetries as function of the maximum total energy quantum number $E_{3\max}$ of the three-body states. For a NCSM calculation of a mid p-shell nucleus in $N_{\max}=8$, corresponding to $E_{3\max}=11$, about $33$ GB are needed to store the necessary 3N matrix elements in single precision exploiting all symmetries \cite{MaAk13}. Moreover, disk-I/O and memory access is nontrivial for these huge sets. In order to extend the NCSM model space to $N_{\max}=12$ or even 14 for mid p-shell nuclei, we have made a first step towards a 'recompute instead of store' strategy in Ref.~\cite{RoLa11}. Instead of precomputing $m$-scheme matrix elements, we only precompute and store the $JT$-coupled matrix elements defined by Eq.~\eqref{eq:JTcoupled_matrixelement}. All the computationally demanding steps of the transformation are still done in the precompute phase. However, as illustrated in Fig.~\ref{fig:memoryneeds}, the storage needed for the $JT$-coupled matrix elements is reduced by up to three orders of magnitude. For an $N_{\max}=8$ p-shell calculation only $0.4\,$GB of storage is needed for the three-body matrix elements in single precision.  

The price to pay for this gain is the on-the-fly decoupling~\eqref{eq:Mscheme_matrixelement} of the three-body matrix elements during the many-body calculation. We have optimized the storage scheme for the $JT$-coupled matrix elements to facilitate a fast and cache-optimized on-the-fly decoupling: we store the values of the matrix elements in a one-dimensional vector. The order and position of the matrix elements is defined via a fixed loop-order for all quantum numbers of the $JT$-coupled matrix elements. The six outer loops are defined by the quantum numbers $\tilde{a}$, $\tilde{b}$, $\tilde{c}$, $\tilde{a}'$, $\tilde{b}'$, $\tilde{c}'$ of the single-particle orbitals, where we exploit antisymmetry and hermeticity. The six inner loops are defined by the coupled quantum numbers $J_{ab}$, $J'_{ab}$, $J$ and $T_{ab}$, $T'_{ab}$, $T$ in this specific order. The three innermost isospin loops run over all 5 possible combinations of the isospin quantum numbers and can be unrolled manually. We do not exploit antisymmetry constraints for matrix elements with identical single-particle orbitals to keep a fixed stride for this inner segment. The angular-momentum loops use the triangular constraints defined through the single-particle quantum numbers. To evaluate a specific $m$-scheme matrix element we jump to the position in the vector defined by the orbital quantum numbers and then evaluate the decoupling loops as a linear sweep over a contiguous segment of the storage vector. Thus, the decoupling operation is very simple and highly cache efficient. This simplicity and its moderate memory footprint makes the decoupling routine an excellent candidate for porting to accelerator cards and first developments along these lines have been successful already~\cite{OrPo13}. The standard implementation of the $JT$-coupled scheme has already been adopted in various many-body methods \cite{MaAk13,CiBa13,HeBo13,HeBi13,BiPi13,BiLa13,RoBi12}.

One could consider to push the boundary further towards recompute in order to save even more memory. Presently we compute and store the $JT$-coupled matrix elements via the transformation~\eqref{eq:JTcoupled_matrixelement} before the many-body calculation. The $T$ coefficients as well as the HOBs, $6j$ and $9j$ symbols that enter Eq.~\eqref{eq:Tcoefficients} are cached for performance reasons. Both, the storage of the resulting $JT$-coupled matrix elements and the caching of the $T$ coefficients requires similar and substantial amounts of memory, as illustrated in Fig.~\ref{fig:memoryneeds}. Therefore, an on-the-fly evaluation of the transformation~\eqref{eq:JTcoupled_matrixelement} using precomputed $T$ coefficients will not reduce the storage needs as compared to the simple decoupling starting from the $JT$-coupled matrix elements as we use it now. In order to save more memory, one would have to evaluate the $T$ coefficients on the fly as well, which results in a significant increase of the computational cost. For present CPU-based architectures the storage of $JT$-coupled matrix elements combined with  the one-the-fly decoupling to the $m$-scheme \eqref{eq:Mscheme_matrixelement} seems to be the optimal compromise.

\section{Similarity Renormalization Group}
\label{sec:srg}

\subsection{Generalities}

Although the interactions from chiral EFT are comparatively soft due to the momentum-space cutoff used to regularize the chiral interactions, it is still difficult to converge NCSM-type calculations beyond the lightest nuclei. Therefore, additional transformations are used to enhance the convergence behavior of the many-body calculation. The two transformation methods that have been successfully used with 3N interactions are the Okubo-Lee-Suzuki (OLS) similarity transformation~\cite{SuLe80,Okub54} and the similarity renormalization group (SRG)~\cite{BoFu07,Wegn94,Wegn00,SzPe00}. The OLS transformation aims at a complete decoupling of a specific many-body model space from the excluded space---as a result the similarity-transformed Hamiltonian depends on basis, model-space size and nucleus. The SRG transformation in its standard formulation~\cite{BoFu07,HeRo07,BoFu10,RoNe10} aims at a generic decoupling of low-momentum or low-energy states from high-lying states and leads to a universal, model-space- and nucleus-independent Hamiltonian. This has significant 
practical advantages, since the same transformed interaction can be used in different many-body approaches, from simple Hartree-Fock-type approaches to coupled-cluster theory and the NCSM.   
Particularly, within the NCSM the fact that the interaction is model-space independent conserves the variational character of the NCSM and enables robust extrapolations to the infinite model space. Therefore, we focus on the SRG transformation in the following.

The basic formulation of the SRG is simple. The Hamiltonian $H$ and all other operators $O$ of interest are subjected to a continuous unitary transformation that can formally be written as
\eq{ \label{eq:srg_transform}
  H_{\alpha} = \adj{U}_\alpha H U_{\alpha} \;,\quad
  O_{\alpha} = \adj{U}_\alpha O U_{\alpha} \;,
}
with a unitary operator $U_{\alpha}$ depending on a continuous parameter $\alpha$, the so-called flow parameter. For $\alpha=0$ we assume $U_{\alpha=0}=1$ and thus $H_{\alpha=0}=H$. Instead of attempting to evaluate the explicit form of the unitariy transformation, we take the derivative of~\eqref{eq:srg_transform} with respect to the flow parameter $\alpha$ and arrive at a first-order differential equation for the evolved Hamiltonian
\eq{ \label{eq:srg_floweq}
  \frac{d}{d\alpha} H_{\alpha} = \comm{\eta_{\alpha}}{H_{\alpha}} \;,
}
with the initial condition $H_{\alpha=0}=H$. The anti-hermitean generator $\eta_{\alpha}$ is connected to the unitary operator $U_{\alpha}$ through another first-order differential equation
\eq{
  \frac{d}{d\alpha} U_{\alpha} = - U_{\alpha} \eta_{\alpha} \;,
}
with initial condition $U_{\alpha=0}=1$. 

At the heart of the SRG is the definition of the generator $\eta_{\alpha}$, which represents the physics encapsulated in the transformation. Once the generator is fixed, the above equations determine the evolved Hamiltonian and all other evolved operators. A variety of SRG generators have been investigated in different physics contexts~\cite{Kehr06,BoFu10}. However, the majority of nuclear structure applications of the SRG use the following definition of the generator
\eq{ \label{eq:srg_generator}
  \eta_{\alpha} = (2\mu)^2\; \comm{T_{\text{int}}}{H_{\alpha}} \;,
}
with the intrinsic kinetic energy $T_{\text{int}} = T - T_{\text{cm}}$ and the reduced nucleon mass $\mu$. Evidently, this generator vanishes if the evolved Hamiltonian and the kinetic energy commute, i.e., if the Hamiltonian is diagonal in the eigenbasis of the kinetic energy operator. This defines a trivial fixed point of the evolution. With increasing flow parameter $\alpha$ the Hamiltonian approaches this fixed point and, thus, it is evolving into a band-diagonal structure with respect to the eigenbasis of the kinetic energy, i.e., momentum eigenstates. For this specific generator it makes sense to associate the flow parameter $\alpha$ with a momentum scale $\lambda_{\text{SRG}}=\alpha^{-1/4}$ as its often done in the literature \cite{BoFu10,JuNa09}. It is important to notice that the generator~\eqref{eq:srg_generator} is not connected to a specific choice of nucleus or basis used in the subsequent many-body calculations. It only reflects the generic goal of decoupling low- and high-momentum components of the model space through a unitary transformation that preserves the complete information of the initial Hamiltonian. 

Owing to its flexibility, the SRG framework can also be adapted to other decoupling scenarios. Considering the $A$-body ground state of a specific nucleus one can design SRG generators that decouple a reference state, e.g., a simple Hartree-Fock determinant representing the nucleus under consideration, from all particle-hole excitations. Once a complete decoupling is achieved, the energy expectation value of the  reference state yields the exact ground-state energy, since, e.g., a full configuration interaction calculation would not admix any particle-hole excitation to this state anymore. In order to handle the SRG evolution in $A$-body space, one can use normal-ordering with respect to the reference state to derive evolution equations for the normal-ordered zero- one- and two-body terms of the Hamiltonian, which are an approximation to the full $A$-body evolution. This defines the so-called in-medium SRG \cite{TsBo11,HeBo13,HeBi13}.

\subsection{Cluster decomposition and basis representation}
\label{subsec:srg_cluster_decom}

All the above equations are general operator relations in an $A$-body Hilbert space or even Fock space. In order to solve them numerically we have to switch to a basis representation in a Hilbert space and we will typically not be able to handle the solution in $A$-body space. We have to rely on solutions of the flow equations in few-nucleon spaces to construct the evolved Hamiltonian. 

This limitation becomes a potential problem since the unitary transformation induces many-body contributions to the evolved operators that go beyond the particle rank of the initial operator. If we assume an initial Hamiltonian containing a two-nucleon interaction, then it is evident from Eqs.~\eqref{eq:srg_floweq} and~\eqref{eq:srg_generator} that an (infinitesimal) step of the flow evolution will induce irreducible operator contributions beyond the two-body level. At any finite flow parameter $\alpha$ the evolved Hamiltonian contains irreducible operator contributions to all particle numbers. This is a simple formal consequence of the fact that the generator $\eta_{\alpha}$ is a two-body operator at least. The same holds for any other evolved operator as well. 

We can decompose the evolved Hamiltonian into contributions to different particle ranks through a cluster expansion~\cite{FeNe98,RoNe10}
\eq{ \label{eq:srg_clusterexp}
  H_{\alpha} 
  = H_{\alpha}^{[1]} + H_{\alpha}^{[2]} + H_{\alpha}^{[3]} + H_{\alpha}^{[4]} + \cdots
}
where $H_{\alpha}^{[k]}$ is an irreducible $k$-body operator that can be formulated in second quantization as
\eqmulti{
  H_{\alpha}^{[k]}
  &= \frac{1}{(k!)^2} \sum_{\alpha_1,...,\alpha_k} \sum_{\beta_1,...,\beta_k} \\
  &\quad  \matrixeas{\alpha_1...\alpha_k}{H_{\alpha}^{[k]}}{\beta_1...\beta_k}\; 
    \adj{a}_{\alpha_1} ... \adj{a}_{\alpha_k} a_{\beta_k} ... a_{\beta_1} \;.
}
The matrix elements of the irreducible $k$-body contribution $H_{\alpha}^{[k]}$ in $k$-body space can be constructed from the matrix elements of the evolved Hamiltonian $H_{\alpha}$ in $k$-body space by simply subtracting the matrix elements of all irreducible operators $H_{\alpha}^{[n]}$ with $n<k$:
\eqmulti{
  &\matrixeas{\alpha_1...\alpha_k}{H_{\alpha}^{[k]}}{\beta_1...\beta_k} = \\
  &\quad= \matrixeas{\alpha_1...\alpha_k}{H_{\alpha}}{\beta_1...\beta_k}
  - \sum_{n=1}^{k-1} \matrixeas{\alpha_1...\alpha_k}{H_{\alpha}^{[n]}}{\beta_1...\beta_k} \;.
}
Thus, if we are able to solve the evolution equations in Hilbert spaces of up to $k$ particles, we can extract all irreducible contributions up to the $k$-body level. Contributions of particle ranks $n$ with $k<n\leq A$ that formally emerge from the unitary transformation in $A$-body space cannot be extracted---we have to truncate the cluster expansion~\eqref{eq:srg_clusterexp}.

The truncation of the cluster expansion at the $k$-body level ($k<A$) formally destroys the unitarity of the transformation in $A$-body space. As long as we preserve unitarity, all eigenvalues of the Hamiltonian in $A$-body space are not changed by the unitary transformation, in particular, all eigenvalues will be independent of the flow parameter $\alpha$. If we discard higher-order terms of the cluster expansion, there is no guarantee that the eigenvalues of the Hamiltonian in $A$-body space are invariant under the transformation. Stated differently, the dependence of the eigenvalues on the flow parameter provides a measure for the impact of the discarded higher-order terms. We will use a systematic flow-parameter variation as a diagnostic for the significance of induced and discarded higher-order contributions later on.

\subsection{Evolution in three-body space}
\label{sec:srg_evolution3b}
 
\begin{figure*}
\includegraphics[width=1.0\textwidth]{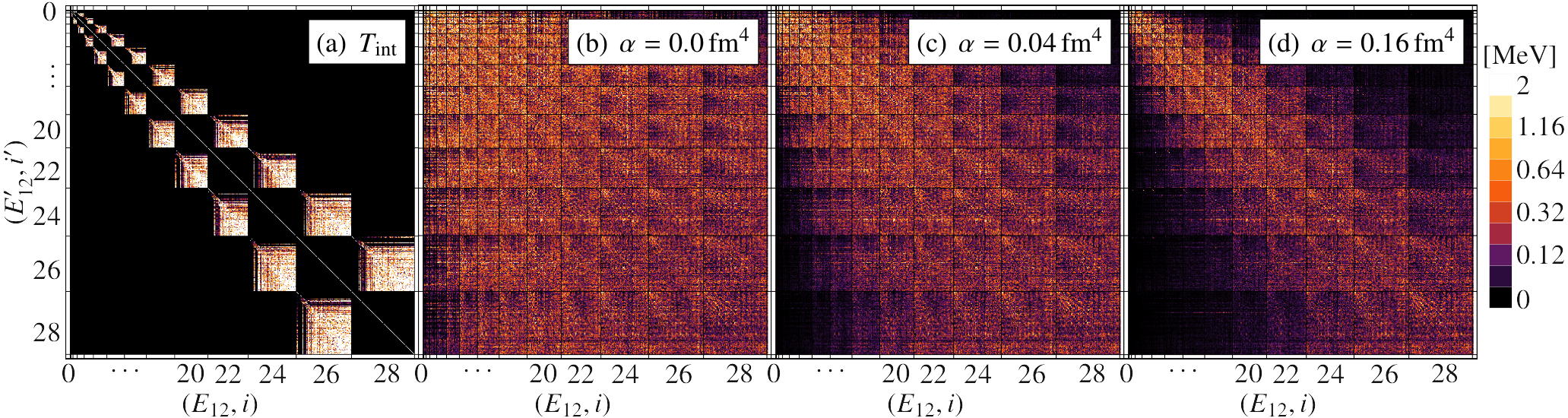}
\caption{(color online) Matrix elements in the antisymmetrized HO Jacobi representation for the triton channel ($J_{12}^{\pi}$, $T_{12}$)=($1/2^+$, $1/2$) for $\hbar\Omega=24\,\text{MeV}$. Plotted are the absolute values of the intrinsic kinetic-energy matrix elements (a) as well as the interaction part of the evolved chiral NN+3N Hamiltonian for flow parameters $\alpha=0\,\text{fm}^4$ (b), $\alpha=0.04\,\text{fm}^4$ (c), and $\alpha=0.16\,\text{fm}^4$ (d). The dark grid lines separate blocks of fixed energy quantum numbers $E_{12}$ and $E'_{12}$. }
\label{fig:me}
\end{figure*}
 
For the numerical solution of the flow equation for the Hamiltonian one can use any computationally convenient basis representation. Two common choices are momentum or HO eigenbases for the relative motion. The center-of-mass degree of freedom can be separated from the beginning, since the Hamiltonian and the generator only act on the relative part of the many-body Hilbert space. Furthermore, in order to exploit the symmetries of the Hamiltonian we use a basis with good total angular momentum, parity, and isospin. 

In two-body space we, thus, use relative LS-coupled momentum or HO eigenstates, i.e., $\ket{q (LS)JT}$ or $\ket{N (LS)JT}$, respectively. The resulting evolution equations in these representations and their solutions are discussed in detail in Refs.~\cite{BoFu10,RoNe10,BoFu07} and we will not repeat the details of the two-body evolution here.

In three-body space we can use the antisymmetrized Jacobi-momentum or Jacobi-HO states introduced in Secs.~\ref{sec:3nme_initial} and~\ref{sec:3nme_jacobiho}, respectively. For reasons of efficiency and technical convenience we use the antisymmetrized Jacobi-HO states to formulate the matrix representation of the evolution equations. Because isospin breaking at the three-body level is expected to have a minor effect, we omit the isospin projection quantum number $M_{T12}$ and use averaged initial three-body matrix elements \cite{Navr07}. Since neither the Hamiltonian nor the generator connect states of different $J_{12}^{\pi}$ and $T_{12}$, the evolution equations decouple for different ($J_{12}^{\pi}$, $T_{12}$) channels. For each channel we obtain, after expansion of the commutators and insertion of two completeness relations,   
\eqmulti{ \label{eq:srg_floweq_3n}
  \frac{d}{d\alpha} & \bra{E_{12}i} \HO_{\alpha} \ket{E'_{12}i'} 
  = (2\mu)^2  \sum_{E''_{12},i''}^{E''_{12} \leq E_{\text{SRG}}} \sum_{E'''_{12},i'''}^{E'''_{12} \leq E_{\text{SRG}}} \big( \\
     &
    \matrixe{E_{12}i}{\TO_{\text{int}}}{E''_{12}i''}\;
    \matrixe{E''_{12}i''}{\HO_{\alpha}}{E'''_{12}i'''}\;
    \matrixe{E'''_{12}i'''}{\HO_{\alpha}}{E'_{12}i'} \\
  -2 & 
  \matrixe{E_{12}i}{\HO_{\alpha}}{E''_{12}i''}\;
    \matrixe{E''_{12}i''}{\TO_{\text{int}}}{E'''_{12}i'''}\;
    \matrixe{E'''_{12}i'''}{\HO_{\alpha}}{E'_{12}i'} \\
   + &
    \matrixe{E_{12}i}{\HO_{\alpha}}{E''_{12}i''}\;
    \matrixe{E''_{12}i''}{\HO_{\alpha}}{E'''_{12}i'''}\;
    \matrixe{E'''_{12}i'''}{\TO_{\text{int}}}{E'_{12}i'} \big) \;,
}
where $\ket{E_{12} i}=\ket{E_{12} i J_{12}^{\pi} T_{12}}_a$ for fixed $J_{12}^{\pi}$ and $T_{12}$. For the completeness relations we of course have to truncate the summation over the infinite three-body basis to a finite model spaces defined by the maximum energy quantum number $E''_{12},E'''_{12} \leq E_{\text{SRG}}$. Note that this flow equation has to be solved also for $E_{12}$ and $E'_{12}$ up to $E_{\text{SRG}}$, since the corresponding matrix elements appear at the right hand side of Eq.~\eqref{eq:srg_floweq_3n}. In practice we reduce the truncation parameter $E_{\text{SRG}}$ with increasing $J_{12}$ since the dimension of the Jacobi-HO basis grows rapidly with $J_{12}$ and since contributions for higher angular momenta have less influence on low-energy nuclear structure observables. We will discuss the details and the impact of this truncation in Sec.~\ref{subsec:role_srg_space}.

Within the finite three-body model space, the numerical problem reduces to a system of coupled linear first-order differential equations for the matrix elements of $\HO_{\alpha}$. The right-hand-side of the flow equation~\eqref{eq:srg_floweq_3n} consists of three-fold matrix products that can be evaluated very efficiently using optimized BLAS matrix multiplications. We use standard solvers with adaptive step size control, e.g., embedded Runge-Kutta methods, to evolve the Hamiltonian up to a given flow parameter $\alpha$. In contrast to early implementations of the SRG evolution in a Jacobi-HO basis~\cite{JuNa09}, the numerical solution of the evolution equations is performed very efficiently---the evolution for the triton channel ($J_{12}^{\pi}$, $T_{12}$)=($1/2^+$, $1/2$) for a typical value of $\alpha$ in a model space with $E_{\text{SRG}}=40$ takes less than one hour on a standard desktop workstation. 

An illustration of the SRG evolution of the three-body matrix elements is presented in Fig.~\ref{fig:me}. We plot the absolute values of the kinetic-energy matrix elements $\braas{E_{12} i J^{\pi}_{12}T_{12}} \TO_{\text{int}} \ketas{E'_{12} i' J^{\pi}_{12}T_{12}}$ and interaction matrix elements $\braas{E_{12} i J^{\pi}_{12}T_{12}} \HO_{\alpha}-\TO_{\text{int}} \ketas{E'_{12} i' J^{\pi}_{12}T_{12}}$ in the antisymmetrized Jacobi-HO representation for the triton channel ($J_{12}^{\pi}$, $T_{12}$)=($1/2^+$, $1/2$) starting from the chiral NN+3N Hamiltonian discussed in Sec.~\ref{sec:prop} for the flow parameters $\alpha=0$, $0.04$, and $0.16\,\text{fm}^4$. The bare interaction shows sizable off-diagonal contributions that are suppressed during the SRG evolution. As a result the Hamiltonian is driven to a band-diagonal form in the Jacobi-HO representation. This is expected from the band-diagonal structure of the intrinsic kinetic energy in the Jacobi-HO basis, which represents a trivial fixed point of the evolution. 

We note that this scheme can be generalized to the evolution in four-body space. The only formal change is the use of an antisymmetrized four-body Jacobi-HO basis. Efforts along these lines are currently under way.

Instead of representing the SRG equations in the Jacobi-HO basis~\eqref{eq:jacobiHO_antisym}, one could also use the Jacobi-momentum representation \eqref{eq:jme3_mom} as shown in Ref.~\cite{Hebe12}. The momentum representation has obvious advantages when aiming at calculations of homogeneous nuclear and neutron matter \cite{HeFu13}. However, for configuration-space nuclear structure calculations build on an underlying HO basis, where one eventually has to provide HO matrix elements, the Jacobi-HO basis has decisive advantages: one can exploit all the benefits of a discrete orthonormal basis, the antisymmetrization of three-body matrix elements is much easier and more efficient, and the typical matrix dimensions to be handled for the numerical solution of the flow-equations are smaller.

A seeming disadvantage of the Jacobi-HO representations is the explicit dependence on the HO oscillator frequency and the need for separate SRG evolutions for each relevant frequency. This and related issued are remedied by using the so-called frequency conversion discussed in the following section.

\subsection{Frequency conversion}
\label{sec:srg_frequencyconv}

Since the evolution equations are solved in the Jacobi-HO basis, we fix the HO frequency $\hbar\Omega$ from the beginning. Thus, in order to perform many-body calculations for different frequencies, we have to perform the SRG evolution for each frequency separately. Depending on the frequency $\hbar\Omega$, the model space used for the SRG evolution spans different momentum or energy ranges. At small frequencies $\hbar\Omega$ the momentum range covered in the SRG model space might not be sufficient to capture the relevant contributions of the initial Hamiltonian. If relevant pieces of the Hamiltonian are discarded already before the SRG evolution due to the $E_{\text{SRG}}$ truncation, then the many-body calculations will exhibit an artificial frequency dependence.   

There is a simple trick to circumvent this problem. We can perform the SRG evolution for a fixed and sufficiently large frequency $\hbar\Omega_{\text{SRG}}$ and afterwards convert the evolved matrix-elements to a smaller frequency $\hbar\Omega$ through a simple basis transformation. For this unitary transformation we need the overlaps of the antisymmetrized Jacobi-HO three-body states $\ket{E_{12} i J_{12}^{\pi} T_{12}}_a$ and $\ket{\tilde{E}_{12} \tilde{i} J_{12}^{\pi} T_{12}}_a$ defined for frequency $\hbar\Omega$ and $\hbar\Omega_{\text{SRG}}$, respectively. These overlaps are given by
\eqmulti{  
  \label{eq:srg_frequencyconv}
   &\braketas{E_{12} i J_{12}^{\pi} T_{12}}{\tilde{E}_{12} \tilde{i} J_{12}^{\pi} T_{12}}= 
   \sum_{N_{1},N_{2}}\,\sum_{\tilde{N}_{1},\tilde{N}_{2}}\, \sum_{\alpha}   \\
   &\quad \quad \delta_{E_{12},2N_1+L_1+2N_2+L_2} C^{i}_{N_1 N_2 \alpha} \,\int d\pi_1\, \pi_{1}^{2}\, R_{N_{1}L_{1}}(\pi_1)\, \tilde{R}_{\tilde{N}_{1}L_{1}}(\pi_1) \\
   &\quad \times \delta_{\tilde{E}_{12},2\tilde{N}_1+L_1+2\tilde{N}_2+L_2} C^{\tilde{i}}_{\tilde{N}_1 \tilde{N}_2 \alpha} \,\int d\pi_2\, \pi_2^{2}\, R_{N_{2}L_{2}}(\pi_2)\, \tilde{R}_{\tilde{N}_{2}L_{2}}(\pi_2) \;,
}
where $R_{NL}(\pi)$ and $\tilde{R}_{\tilde{N}L}(\pi)$ are the radial HO wave functions associated with frequency $\hbar\Omega$ and $\hbar\Omega_{\text{SRG}}$, respectively, and $C^{i}_{N_1 N_2 \alpha}$ are the CFPs.

Obviously, this basis transformation also needs to be truncated to a finite model space. However, as the frequency conversion is performed after the SRG evolution the Hamiltonian already has a band-diagonal structure and the low- and high-momentum basis states are decoupled. The frequency transformation, described by the matrix of overlaps~\eqref{eq:srg_frequencyconv}, which itself has a band-diagonal structure, will only mix matrix elements from a limited region. The low-energy sector of the Jacobi-HO matrix-elements that enters the many-body calculation later on is thus not affected by the truncation of the model space during the frequency conversion.

We will investigate the effect of the frequency conversion and the impact of the SRG model-space truncation in actual many-body calculations in Sec.~\ref{subsec:role_srg_space}.

\section{Importance-Truncated No-Core Shell Model}
\label{sec:itncsm}

\subsection{Generalities}

The no-core shell model (NCSM) is one of the standard ab initio methods in nuclear structure theory~\cite{BaNa13,NaQu09}. It is conceptually simple and very flexible: the eigenvalue problem of the Hamiltonian is solved numerically in a finite many-body basis representation yielding the energy eigenvalues and eigenstates, which give access to all observables. It obeys the variational principle and various extrapolation techniques to the infinite Hilbert space can be used \cite{MaVa09,CoAv12,FuHa12,MoEk13}. From the point of view of general configuration interaction (CI) approaches, the NCSM is based on two defining elements: (i) the many-body basis is build from HO eigenstates formulated either in single-particle or in Jacobi coordinates, and (ii) the many-body model space is truncated with respect to the unperturbed excitation energy $N_{\max}\hbar\Omega$ of the HO many-body basis states. 

One of the specific advantages resulting from (i) and (ii) is the equivalence of the single-particle and the Jacobi coordinate formulation of the NCSM~\cite{NaKa00}. As a practical consequence an NCSM calculation with a translational invariant Hamiltonian using a basis of Slater determinants of single-particle HO states leads to eigenstates that factorize exactly into a center-of-mass and a relative component---this is not the case for other choices of the single-particle basis or many-body truncations. Another advantage of the many-body truncation (ii) as compared to truncations with respect to the single-particle basis is its computational efficiency. The many-body basis dimension needed to approach the exact result to a given accuracy is much smaller for the $N_{\max}$ truncation than for a truncation of the maximum single-particle energy quantum number $e_{\max}$. This indicates that the truncation guided by the many-body energy is physically more adequate than a truncation 
based on single-particle energies.         

Still, the basic limitation of the NCSM results from the combinatorial growth of the many-body basis dimension with particle number $A$ and truncation parameter $N_{\max}$. In order to slow down this growth we have proposed an additional importance truncation of the NCSM model space in Refs.~\cite{RoNa07,Roth09}. The basic idea is to selectively remove basis states from the full NCSM model space using an adaptive, state-specific, and physics-guided truncation criterion. 

Assume we target a small number of low-energy eigenstates $\ket{\Psi^{(m)}}$ for $m=1,...,M$ in an NCSM calculation for a specific $N_{\max}$. The full NCSM calculation would yield eigenvectors representing the amplitudes $C_{\nu}^{(m)}$ for the expansion of the target eigenstates in terms of the many-body basis states $\ket{\Phi_{\nu}}$:   
\eq{
  \ket{\Psi^{(m)}} = \sum_{\nu} C_{\nu}^{(m)} \ket{\Phi_{\nu}} \; .
}
Many of the amplitudes will have very small or vanishing values, i.e., the corresponding basis states do not contribute significantly to the target states. If these amplitudes were known a priori, we could have reduced the basis dimension significantly by discarding those basis states and would still obtain a good variational approximation of the target states.

In order to estimate the amplitudes a priori, we use initial approximations of the target states, so-called reference states $\ket{\Psi^{(m)}_{\text{ref}}}$, that are typically determined from a previous NCSM calculation in a smaller model space $\mathcal{M}_{\text{ref}}$ 
\eq{
  \ket{\Psi^{(m)}_{\text{ref}}} = \sum_{\nu \in \mathcal{M}_{\text{ref}}} C_{\text{ref},\nu}^{(m)} \ket{\Phi_{\nu}} \; .
}
These reference states carry information about the physical properties of the target eigenstates. Guided by first-order multiconfigurational perturbation theory we estimate the amplitudes of the individual basis states $\ket{\Phi_{\nu}}\notin \mathcal{M}_{\text{ref}}$ in the expansion of the target eigenstate. This first-order perturbative correction for the amplitudes defines the so-called importance measure
\eq{\label{eq:Importance_Measure}
  \kappa_{\nu}^{(m)} = -\frac{ \matrixe{\Phi_{\nu}}{\HO}{\Psi_{\text{ref}}^{(m)}}}{\Delta \epsilon_{\nu}} \;,
}
where $\HO$ is the full Hamiltonian of the NCSM calculation and $\Delta \epsilon_{\nu}$ is an energy denominator which is taken to be the unperturbed HO excitation energy of the basis state $\ket{\Phi_{\nu}}$~\cite{Roth09,RoNa07}. 

The importance measure combines information about the properties of the target states, carried by the reference states, about the many-body basis, and about the Hamiltonian and is the basis for the definition of a state-dependent adaptive truncation of the model space, the so-called importance truncation (IT). We define the importance-truncated model-space $\mathcal{M}_{\text{IT}}(\kappa_{\min})$ spanned by all states of the reference space $\mathcal{M}_{\text{ref}}$ plus all basis states $\ket{\Phi_{\nu}}\notin \mathcal{M}_{\text{ref}}$ with importance measure $| \kappa_{\nu}^{(m)} | \geq \kappa_{\min}$ for at least one $m \in \{1,...,M\}$. The importance threshold $\kappa_{\min}$ provides an additional truncation parameter, which will be varied later on to probe the contribution of the discarded basis states. Note, that the importance measure~\eqref{eq:Importance_Measure} is based on the first-order perturbative correction to the states, not on the perturbative correction to the energies. It is, therefore, not biased to an optimal description of energies, but aims at an optimal description of the states and, thus, of all observables.

\subsection{Construction of the IT-NCSM model space}

There are different ways to embed the importance truncation into general CI-type calculations~\cite{Roth09}. In the context of the $N_{\max}$-truncated model space of the NCSM, a sequential scheme has proven to be most efficient. In practice we are always interested in NCSM calculations for a sequence of $N_{\max}$ values in order to assess the convergence with respect to the model-space size or to perform extrapolations to the infinite space. 

The increase $N_{\max} \to N_{\max}+2$ to the next-larger same-parity NCSM space can be elegantly combined with the importance-truncation procedure: We use the eigenstates $\ket{\Psi^{(m)}}$ obtained in the $N_{\max}$-space to define the reference state for the construction of the importance-truncated $N_{\max}+2$ space. Generally, we do not keep the full eigenstate as reference state, but introduce a second threshold parameter, the so-called reference threshold $C_{\min}$. The reference space $\mathcal{M}_{\text{ref}}$ is spanned by all basis states of the $N_{\max}$ space with amplitudes $| C_{\nu}^{(m)} | \geq C_{\min}$ for at least one $m \in \{1,...,M\}$. The reference states $\ket{\Psi^{(m)}_{\text{ref}}}$ are the normalized projections of the eigenstates $\ket{\Psi^{(m)}}$ onto the reference space $\mathcal{M}_{\text{ref}}$. These reference states are used to construct the importance-truncated $N_{\max}+2$ space. Note that during the importance selection, all basis states from the reference space $\mathcal{M}_{\text{ref}}$ are retained and all states from the full $N_{\max}+2$ model space that are not in $\mathcal{M}_{\text{ref}}$ are probed. 

In order to efficiently generate a sequence of importance thresholds $\kappa_{\min}$ for an a posteriori extrapolation, we start with constructing the importance-truncated space for the smallest $\kappa_{\min}$ in the sequence, corresponding to the largest model space. The model spaces for the larger importance thresholds $\kappa_{\min}$ are then obtained by filtering out basis states based on the previously determined $\kappa_{\nu}$. Thus, the time-consuming importance selection is performed only once for each $\kappa_{\min}$-set.

For accessing a sequence of $N_{\max}$ spaces, we start with full NCSM calculations up to a convenient $N_{\max}$, typically $N_{\max}=4$ or $6$. Beyond that we use the sequential importance update described above to increase $N_{\max}$ in steps of $2$. An important formal property of this scheme is that in the limit of vanishing thresholds $(\kappa_{\min},C_{\min})\to 0$ we will recover the sequence of full $N_{\max}$-spaces. This is the basis for the extrapolation procedures discussed in the next section. 

Further details on the algorithm we developed for the importance truncation can be found in Ref.~\cite{Roth09}. A slightly different implementation of this importance-truncation technique was recently presented in Ref.~\cite{KrJu13}.

\subsection{Threshold extrapolation \& uncertainty quantification}

The importance truncation is constructed to retain only the physically important states of the many-body basis, where the distinction between important and unimportant states is controlled by the importance threshold $\kappa_{\min}$. Still, the discarded states will have a quantitative effect on the many-body observables we aim to compute, and we have to try to recover their contribution to arrive at an accurate result. 

The simplest way to estimate the effect of the discarded basis states on the energy is through a second-order perturbative correction. During the construction of the importance-truncated space, we can evaluate a second-order estimate for the energy contribution of the basis state $\ket{\Phi_{\nu}}$ through
\eq{
  \xi^{(m)}_{\nu} = -\frac{ | \matrixe{\Phi_{\nu}}{\HO}{\Psi_{\text{ref}}^{(m)}} |^2}{\Delta \epsilon_{\nu}}
}
at no additional cost. Whenever basis states are discarded, i.e., if $|\kappa_{\nu}| \leq \kappa_{\min}$, we accumulate their second-order energy contributions $\xi^{(m)}_{\nu}$ in an estimate for the energy contribution of the excluded states $\Delta^{(m)}_{\text{excl}}(\kappa_{\min})$. This correction can be added a posteriori to the energy eigenvalues $E^{(m)}_{\text{eval}}(\kappa_{\min})$ obtained in the importance-truncated space. Unfortunately, this correction is not easily available for other observables than the energy. 

Another way to assess the contributions of excluded basis states is through a variation of the importance threshold $\kappa_{\min}$. The energy eigenvalues $E^{(m)}_{\text{eval}}(\kappa_{\min})$ are smooth functions of $\kappa_{\min}$ and they decrease monotonically with decreasing $\kappa_{\min}$ as dictated by the variational principle. Other observables, which are evaluated via expectation values or matrix elements with the energy eigenstates from an importance-truncated spaces, also exhibit a smooth, but not necessarily monotonic dependence on the importance threshold $\kappa_{\min}$. In addition it is guaranteed that all observables will approach their values in the full NCSM space in the limit of vanishing thresholds $(C_{\min},\kappa_{\min}) \to 0$. Together, these properties motivate the a posteriori extrapolation of the observables to $(C_{\min},\kappa_{\min}) \to 0$ in order to recover the contributions of excluded configurations and thus the full NCSM 
result up to uncertainties of the 
extrapolation. 

In practical calculations we choose the reference threshold $C_{\min}$ small enough so that it does not affect the results. The remaining extrapolation of the importance threshold $\kappa_{\min}\to 0$ is performed using simple polynomials $P_n(\kappa_{\min})$ of different orders $n$. It is important to note that this extrapolation is the only source of systematic uncertainties in an IT-NCSM calculation compared to the full NCSM result. Therefore, it is important to quantify and control this uncertainty. We do this on a case-by-case basis for each observable and state as a routine part of the many-body calculation. Starting from a set of IT-NCSM calculations for a sequence of thresholds $\kappa_{\min}$, typically eight values in the range from $3\times 10^{-5}$ to $10\times 10^{-5}$, at fixed $C_{\min}$, typically $2\times 10^{-4}$, we construct a family of fits. The fit to the full data set with a polynomial $P_n(\kappa_{\min})$, typically of order $n=3$, provides the $\kappa_{\min}\to0$ extrapolated value 
of the observable. Additional extrapolations with orders $n+1$ and $n-1$ to the full data set, as well as extrapolations of order $n$ with lowest and the lowest two $\kappa_{\min}$-results dropped define an uncertainty band for the extrapolation. The span of this uncertainty band for $\kappa_{\min}\to0$ provides an individual measure of the systematic uncertainty for each threshold-extrapolated observable extracted from an IT-NCSM calculation. The whole analysis can be repeated for a different value of the reference threshold $C_{\min}$ in order to confirm that it does not affect the threshold-extrapolated observables.    

For the description of energies, we can combine the threshold extrapolation with information obtained for the second-order energy correction due to the excluded states. We make use of the trivial fact that the energy correction $\Delta^{(m)}_{\text{excl}}(\kappa_{\min})$ vanishes in the limit $\kappa_{\min}\to0$. Therefore, we can construct a family of improved energy curves 
\eq{
  E^{(m)}_{\lambda}(\kappa_{\min}) = E^{(m)}_{\text{eval}}(\kappa_{\min}) + \lambda\; \Delta^{(m)}_{\text{excl}}(\kappa_{\min})
}
with an auxiliary control parameter $\lambda$, that are guaranteed to approach the same value $E^{(m)}_{\lambda}(\kappa_{\min}) \to E^{(m)}_0$ in the limit $\kappa_{\min}\to 0$ independent of $\lambda$. Therefore, with a given set of $\lambda$-parameters we can perform a simultaneous fit of a set of polynomials to each of the $E^{(m)}_{\lambda}(\kappa_{\min})$-curves under the constraint that $E^{(m)}_{\lambda}(\kappa_{\min}=0) = E^{(m)}_0$ for all $\lambda$. Since the different $E^{(m)}_{\lambda}(\kappa_{\min})$ curves typically approach the common value $E^{(m)}_0$ from both directions we achieve a substantial stabilization of the extrapolation. 

We can use the same method to estimate the uncertainties as in the simple extrapolation. In addition to varying the order of the fit polynomials by $\pm 1$ and omitting the one or two lowest-$\kappa_{\min}$ points, we vary the set of $\lambda$ values used for the simultaneous constrained extrapolation by omitting the largest or smallest $\lambda$. We again arrive at an error band and an intrinsic and state-specific estimate for the systematic uncertainty due to the importance truncation and extrapolation. 

\begin{figure}
\includegraphics[width=0.91\columnwidth]{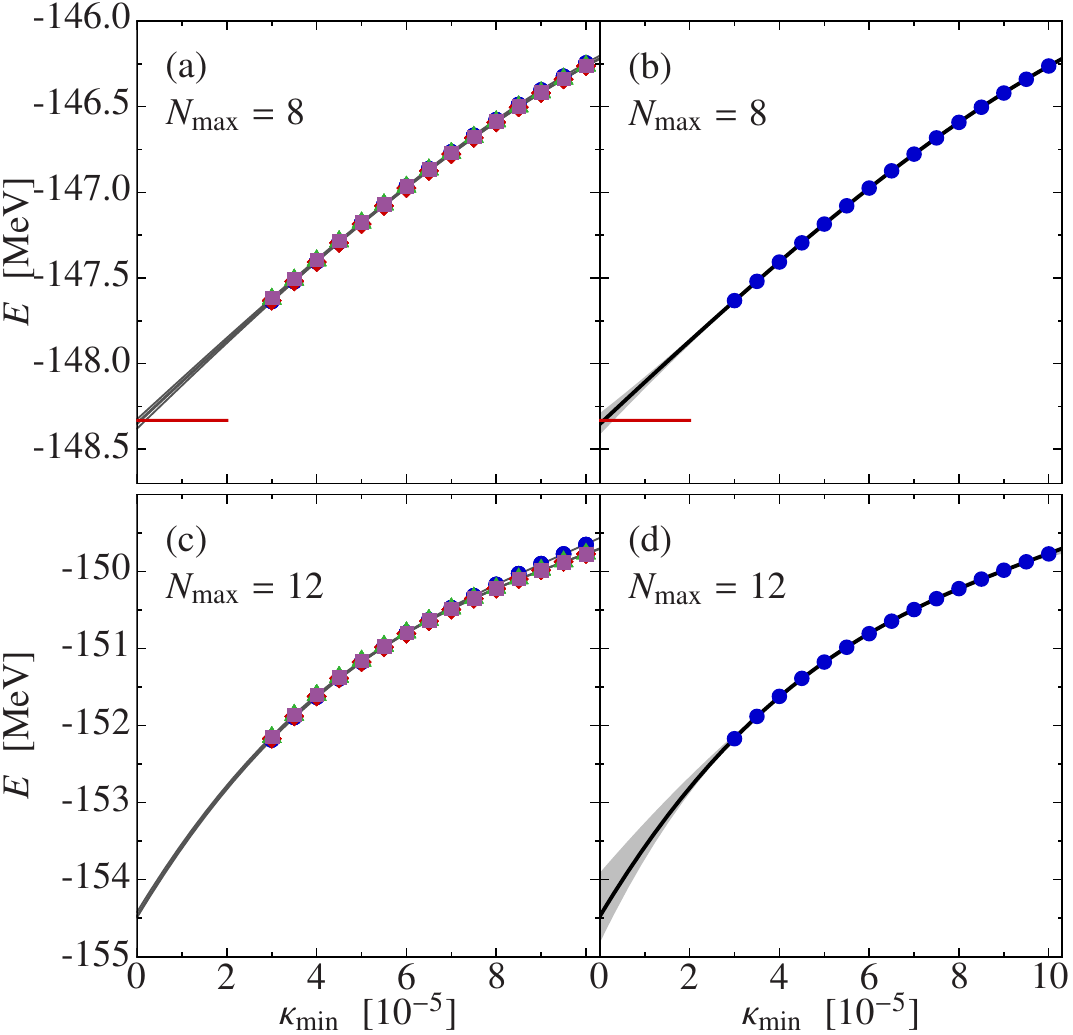}
\caption{(color online) Threshold dependence and extrapolation for the ground-state energy of \elem{O}{16} with the chiral NN interaction evolved at the two-body level to $\alpha=0.04\,\text{fm}^4$ at $\hbar\Omega=20\,\text{MeV}$. Panel (a) and (c) show the $\kappa_{\min}$-dependence of the ground-state energy for different reference thresholds $C_{\min}=1\times10^{-4}$ (\symbolcircle[FGBlue]), $2\times10^{-4}$ (\symboldiamond[FGRed]), $3\times10^{-4}$ (\symboltriangle[FGGreen]), and $5\times10^{-4}$ (\symbolbox[FGViolet]) . Panels (b) and (d) illustrate the simple threshold extrapolation for $C_{\min}=2\times10^{-4}$ using a third-order polynomial with uncertainty bands derived from the extrapolation protocol described in the text. Red bars mark the full NCSM results obtained with the \textsc{Antoine} code \cite{CaNo99}.}
\label{fig:itncsm_O16_thresholdextrapol1}
\end{figure}
\begin{figure}
\includegraphics[width=0.75\columnwidth]{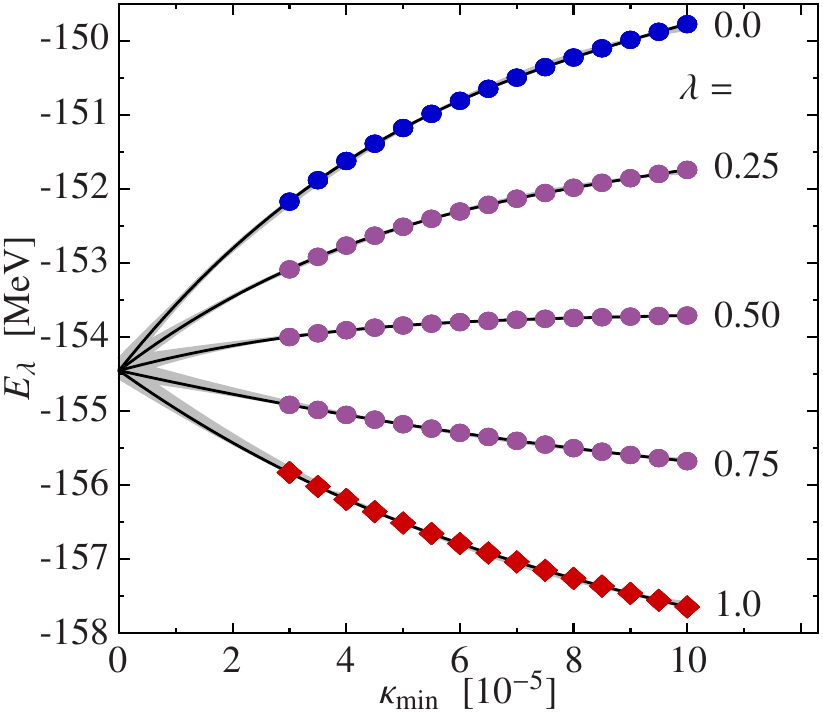}
\caption{(color online) Illustration of the constrained simultaneous extrapolation for the ground-state energy of \elem{O}{16} with the chiral NN interaction evolved at the two-body level to $\alpha=0.04\,\text{fm}^4$ at $\hbar\Omega=20\,\text{MeV}$ for $N_{\max}=12$ and $C_{\min}=2\times10^{-4}$.}
\label{fig:itncsm_O16_thresholdextrapol2}
\end{figure}

A first set of examples for the threshold extrapolation are shown in Figs.~\ref{fig:itncsm_O16_thresholdextrapol1} and~\ref{fig:itncsm_O16_thresholdextrapol2}. For this demonstration we use the chiral NN interaction at N${}^{3}$LO by Entem \& Machleidt \cite{EnMa03} with an SRG evolution at the two-body level, later referred to as NN-only Hamiltonian. We consider the ground-state energies of \elem{O}{16} obtained for $N_{\max}=8$ and $N_{\max}=12$. For these IT-NCSM calculations we start with a full NCSM run for $N_{\max}=4$ and then use the sequential update scheme to increase the model-space size in steps of $2$ using the lowest eigenstate from the previous model-space as reference state. Figures~\ref{fig:itncsm_O16_thresholdextrapol1}(a) and (c) illustrate the effect of the reference threshold $C_{\min}$. The data sets for different $C_{\min}$ were obtained by computing the whole $N_{\max}$-sequence with different but fixed values of $C_{\min}$. We find virtually no dependence on the reference thresholds throughout the whole range from $C_{\min}=1\times10^{-4}$ to $5\times10^{-4}$. 

Figures~\ref{fig:itncsm_O16_thresholdextrapol1}(b) and (d) illustrate the simple threshold extrapolation with the error bands resulting from the protocol discussed above. For $N_{\max}=8$ we have the exact NCSM result for the ground-state energy obtained with the \textsc{Antoine}-code~\cite{CaNo99} for comparison. The simple extrapolation reproduces the exact value within the estimated extrapolation uncertainty, which is very small. For $N_{\max}=12$, where a full NCSM calculation is not possible anymore, the extrapolation uncertainties increase, but are still well under control. 

In order to reduce the uncertainties for the $N_{\max}=12$ extrapolation, we can adopt the simultaneous constained extrapolation scheme making use of the perturbative corrections for the excluded configurations. This is illustrated in Fig.~\ref{fig:itncsm_O16_thresholdextrapol2} for the $N_{\max}=12$ calculation. The set of auxiliary $\lambda$-parameters is chosen such that the $E_{\lambda}(\kappa_{\min})$ curves exhibit an approximately symmetrical approach, which stabilized the extrapolation to $\kappa_{\min}\to0$ significantly and also reduces the uncertainty band. In practical applications, we use the simple extrapolation as long as the uncertainties are in an acceptable range and switch to the constrained extrapolation only if necessary to obtain a stable extrapolation.
 
\begin{figure}
\includegraphics[width=\columnwidth]{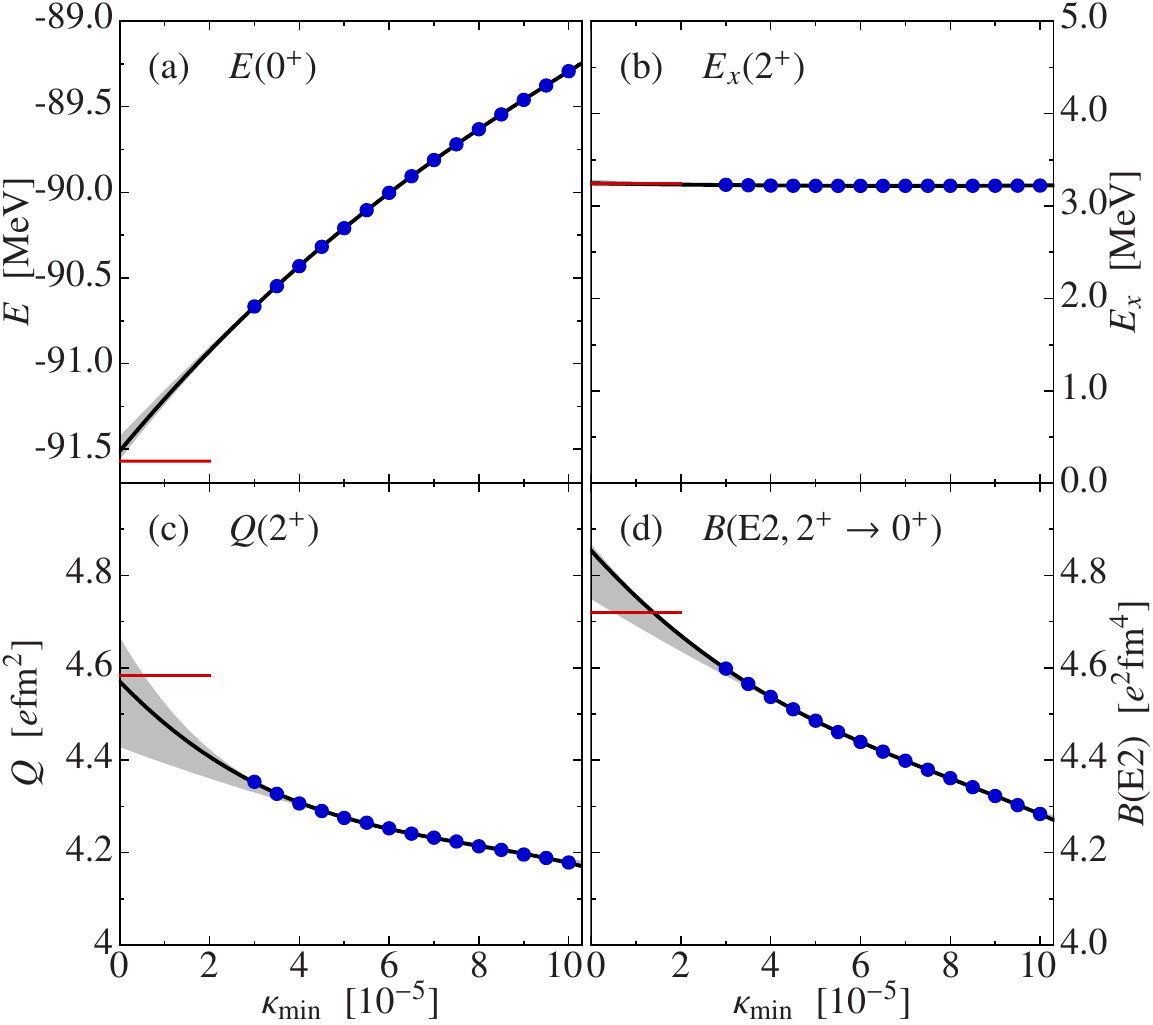}
\caption{(color online) Threshold dependence and extrapolation for (a) the ground-state energy of \elem{C}{12} at $N_{\max}=8$, (b) the excitation energy of the first excited $2^+$ state, (c) the quadrupole moment of the $2^+$ state, and (d) the $B(E2)$ transition strength from the $2^+$ to the ground state. We use the evolved chiral NN interaction at $\alpha=0.04\,\text{fm}^4$ and $\hbar\Omega=20\,\text{MeV}$ with $C_{\min}=2\times 10^{-4}$. The black lines show third-order threshold extrapolations with the gray uncertainty bands obtained from the extrapolation protocol. Red bars mark the full NCSM results obtained with the \textsc{Antoine} code \cite{CaNo99}.}
\label{fig:itncsm_C12_thresholdextrapol}
\end{figure}

As a second set of examples for the threshold extrapolation we consider the ground and the first excited $2^+$ state in \elem{C}{12}. In Fig.~\ref{fig:itncsm_C12_thresholdextrapol} we show the threshold dependence of the ground-state energy, the excitation energy of the first $2^+$ state, the quadrupole moment of the first $2^+$ and the $B(\text{E2})$ transition strength from the $2^+$ to the ground state. In all cases we use the simple threshold-extrapolation scheme with a third-order polynomial and the uncertainty estimation discussed above. Whereas the ground-state energy shows the same $\kappa_{\min}$ dependence as the previous cases, the excitation energy is practically independent of the threshold $\kappa_{\min}$, i.e., the threshold dependence of the absolute energies of both states is very similar and cancels when considering their difference. This enables us to extract excitation energies with much smaller uncertainties than absolute energies. 

For other spectroscopic observables, in particular electric quadrupole moments and transition strengths, the threshold extrapolation is more difficult. These observables are very sensitive to the long-range behavior of the wave functions, which is typically determined by the small components of the HO basis expansion. Therefore, the importance truncation affects these quantities more severely than the energies. Furthermore, unlike the energy, these observables are not protected by the variational principle and can exhibit a more complicated non-monotonous threshold dependence. Together, these properties lead to larger uncertainties in the threshold extrapolation, which are evident from the examples shown in Fig.~\ref{fig:itncsm_C12_thresholdextrapol}(c) and (d). Nevertheless, for spectroscopic observables such as magnetic dipole moments and transitions that do not exhibit a pronounced dependence on the long-range behavior of the wave function, the threshold extrapolations are simple and accurate. 

In addition to the systematic uncertainties resulting from the importance truncation and threshold extrapolation, the IT-NCSM faces the same uncertainties due to the model-space truncation in terms of $N_{\max}$ as the standard NCSM. We will come back to these model-space extrapolations in Sec.~\ref{sec:comp}.

\section{Properties of SRG-evolved Hamiltonians}
\label{sec:prop}

Using the IT-NCSM we now assess the properties of the SRG-evolved Hamiltonians relevant for the application in many-body calculations.

We start from the chiral NN interaction at N${}^3$LO by Entem and Machleidt \cite{EnMa03} and the chiral 3N interaction at N${}^2$LO in the local formulation by Navr\'atil \cite{Navr07}. If not stated otherwise, the 3N interaction uses a cutoff $\Lambda_{\text{3N}}=500\,\text{MeV}/c$ and low-energy constants $c_{D}$ and $c_{E}$ are fitted to the ground-state energy of $A=3$ systems and the $\beta$-decay half-life of \elem{H}{3} \cite{GaQu09}. The initial 3N matrix elements in the antisymmetrized Jacobi-HO basis are obtained directly from Petr Navr\'atil's \textsc{ManyEff} code \cite{NaKa00}. 

We perform the SRG evolution of the NN interaction in two-body space using momentum-space partial-wave matrix elements on a sufficiently fine and large momentum grid. The three-body part of the evolved Hamiltonian is determined from an evolution in the three-body Jacobi-HO basis with a consistent subtraction of the two-body part evolved in a HO basis of compatible size. Depending on which of the three-body contributions are considered, we define the following Hamiltonians \cite{JuNa09,RoLa11}: the NN-only Hamiltonian only uses the initial chiral NN interaction and keeps only two-body contributions throughout the SRG evolution. The NN+3N-induced Hamiltonian starts from the initial NN interaction and keeps the SRG-evolved two- and three-body terms. The NN+3N-full Hamiltonian starts from an initial NN+3N Hamiltonian and again keeps SRG-evolved two- and three-body terms. In all Hamiltonians induced four-body and multi-nucleon contributions are omitted and we use the variation of the SRG flow-parameter to assess the effect of these terms.

\subsection{Role of the SRG model space}
\label{subsec:role_srg_space}

As a first technical aspect we discuss the details and investigate the impact of the truncation of the SRG model space mentioned in Sec.~\ref{sec:srg_evolution3b}. In Eq.~\eqref{eq:srg_floweq_3n} we have introduced the truncation parameter $E_{\text{SRG}}$ for the three-body Jacobi-HO basis used for solving the SRG evolution equations. For fixed  $E_{\text{SRG}}$ the basis dimension of a ($J_{12}^{\pi}$, $T_{12}$) channel grows rapidly with increasing $J_{12}$. At the same time, channels with large $J_{12}$ are of lesser importance for the description of low-energy properties of light nuclei. Therefore, we introduce a $J_{12}$-dependent truncation parameter $E_{\text{SRG}}(J_{12})$ which decreases with increasing $J_{12}$. 

\begin{figure}
\includegraphics[width=0.75\columnwidth]{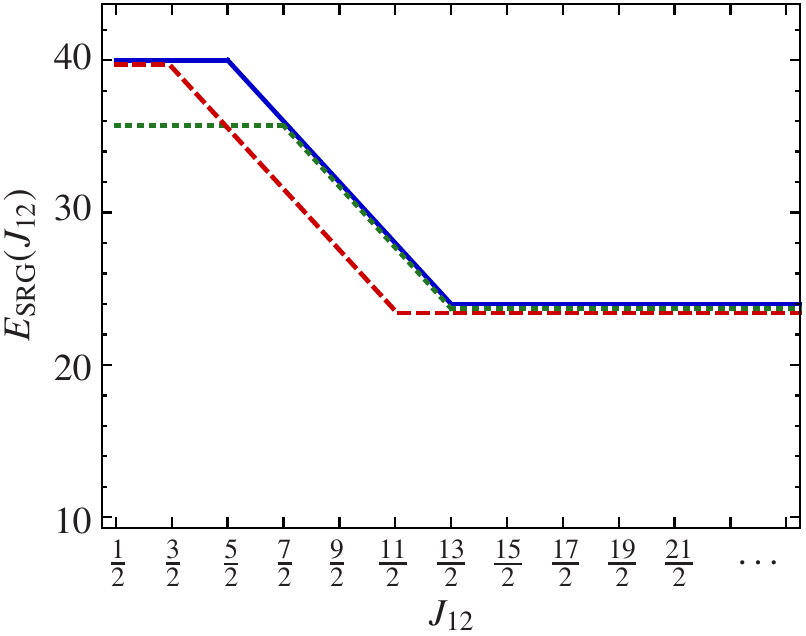}
\caption{(color online) Schematic presentation of the SRG model-space truncation parameter $E_{\text{SRG}}$ depending on the angular momentum $J_{12}$. Plotted are ramp A (blue solid line), ramp B (red dashed line), ramp C (green dotted line).}
\label{fig:prop_srgramp}
\end{figure}


Figure~\ref{fig:prop_srgramp} illustrates three specific choices for $E_{\text{SRG}}(J_{12})$, the so-called ramps, that we adopt in the following. Ramp A defines our default choice for the SRG model space: all three-body channels up to $J_{12}=5/2$ use $E_{\text{SRG}}=40$, beyond that we reduce $E_{\text{SRG}}$ in steps of $4$ until $J_{12}=13/2$ and beyond we use $E_{\text{SRG}}=24$. Ramps B and C are used to study the effect of the $E_{\text{SRG}}$ truncation on many-body observables---the former starts reducing $E_{\text{SRG}}$ already for $J_{12}=5/2$ and the latter uses $E_{\text{SRG}}=36$ for $J_{12}\leq 7/2$. In a series of previous publications~\cite{RoLa11,RoBi12,BiLa13,HeBo13,HeBi13} we have always used ramp A, whereas other groups typically choose other schemes to reduce $E_{\text{SRG}}$ with increasing $J_{12}$~\cite{JuNa09,JuNa11,JuMa13}. 

\begin{figure}
\includegraphics[width=\columnwidth]{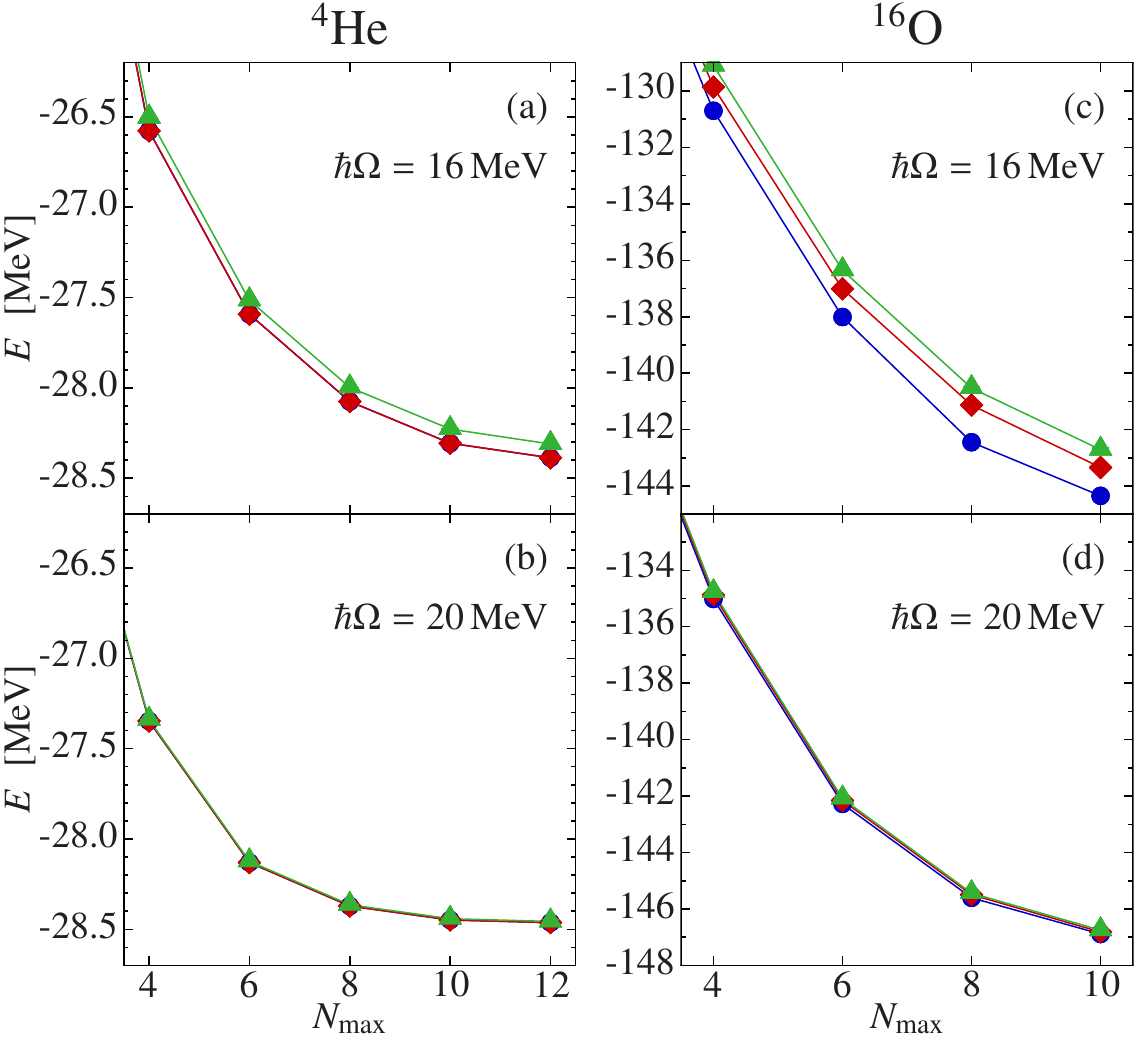}
\caption{(color online) Ground-state energy of \elem{He}{4} and \elem{O}{16} with the NN+3N-full interaction for $\hbar\Omega=16,\,20\,\text{MeV}$ and $\alpha=0.08\,\text{fm}^{4}$ as function of $N_{\text{max}}$. The three curves correspond to the used SRG model space truncations defined by ramp A (\symbolcircle[FGBlue]), ramp B (\symboldiamond[FGRed]), and ramp C (\symboltriangle[FGGreen]).  }
\label{fig:prop_srgramp_gsenergy}
\end{figure}

We first analyze the dependence of IT-NCSM ground-state energies of \elem{He}{4} and \elem{O}{16} on the SRG model space. In Fig.~\ref{fig:prop_srgramp_gsenergy} we show the $N_{\max}$-dependence of the ground-state energies obtained with the NN+3N-full Hamiltonian for $\alpha=0.08\,\text{fm}^4$ for two different HO frequencies. For $\hbar\Omega=20\,\text{MeV}$, depicted in Fig.~\ref{fig:prop_srgramp_gsenergy} (b) and (d), we find that the energies of both nuclei are independent of the choice of the SRG model space, i.e., the results obtained with all three ramps are on top of each other. However, when going to the lower frequency $\hbar\Omega=16\,\text{MeV}$, as shown in Fig.~\ref{fig:prop_srgramp_gsenergy} (a) and (c), we observe a sizable dependence of the ground-state energies on the SRG model-space. For \elem{He}{4} the ramps A and B provide the same results but ramp C gives $0.4\%$ less binding. For \elem{O}{16} the results for ramps B and C both differ from ramp A on a scale of up to $1.5\%$. Together, this indicates that for $\hbar\Omega=16\,\text{MeV}$ the $E_{\text{SRG}}$ truncation of low-$J_{12}$ channels becomes visible and that for heavier nuclei also the ramping-down of $E_{\text{SRG}}$ with increasing $J_{12}$ affects the absolute energies. We have confirmed this trend already in coupled-cluster calculations extending into the mass $A\sim50$ region~\cite{RoBi12,BiLa13}.

\begin{figure}
\includegraphics[width=\columnwidth]{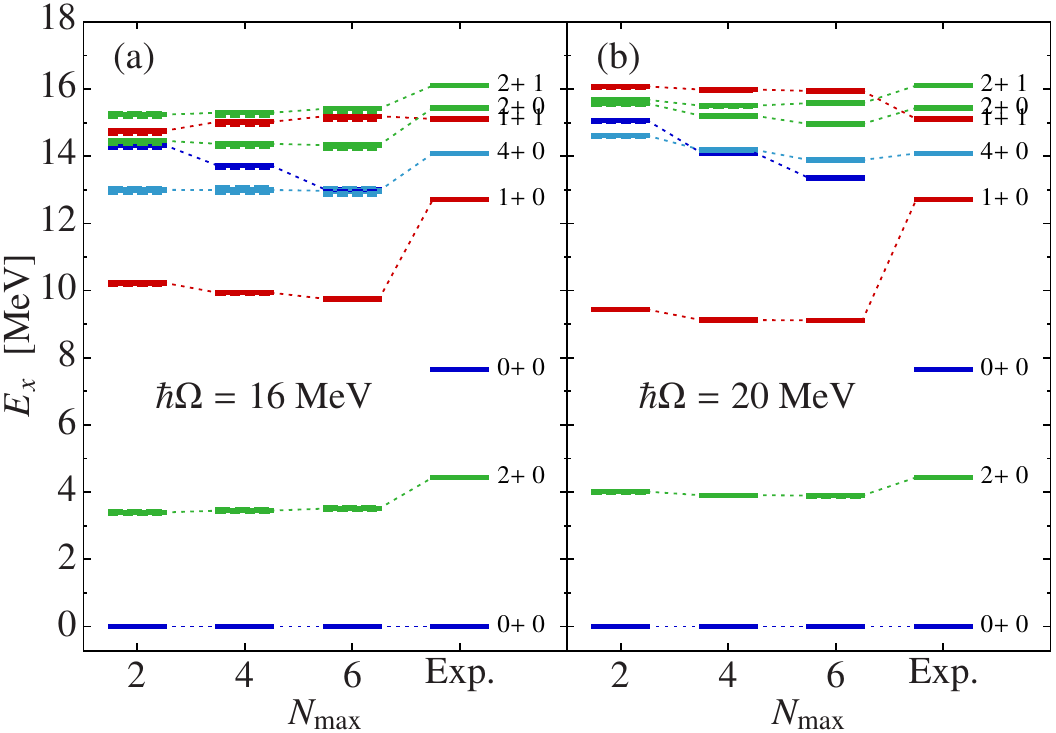}
\caption{(color online) Excitation spectrum of \elem{C}{12} with the NN+3N-full Hamiltonian for $\alpha=0.08\,\text{fm}^4$ and HO frequencies $\hbar\Omega=16\,\text{MeV}$ (a) and $20\,\text{MeV}$ (b). Three sets of calculations are shown (almost always on top of each other) using SRG model-space truncations defined by ramp A (solid bars), ramp B (dashed bars), and ramp C (dotted bars).} 
\label{fig:prop_srgramp_excitation}
\end{figure}

The effect of the SRG model space on excitation energies is much weaker, as illustrated in Fig.~\ref{fig:prop_srgramp_excitation} for the excitation spectrum of \elem{C}{12}. Even for frequency $\hbar\Omega=16\,\text{MeV}$ the excitation spectra obtained with the three different ramps are essentially the same. Thus, the parts of the Hamiltonian that are not captured in the SRG-model space only cause a shift of the whole spectrum without influencing details of its structure.

\begin{figure}
\includegraphics[width=\columnwidth]{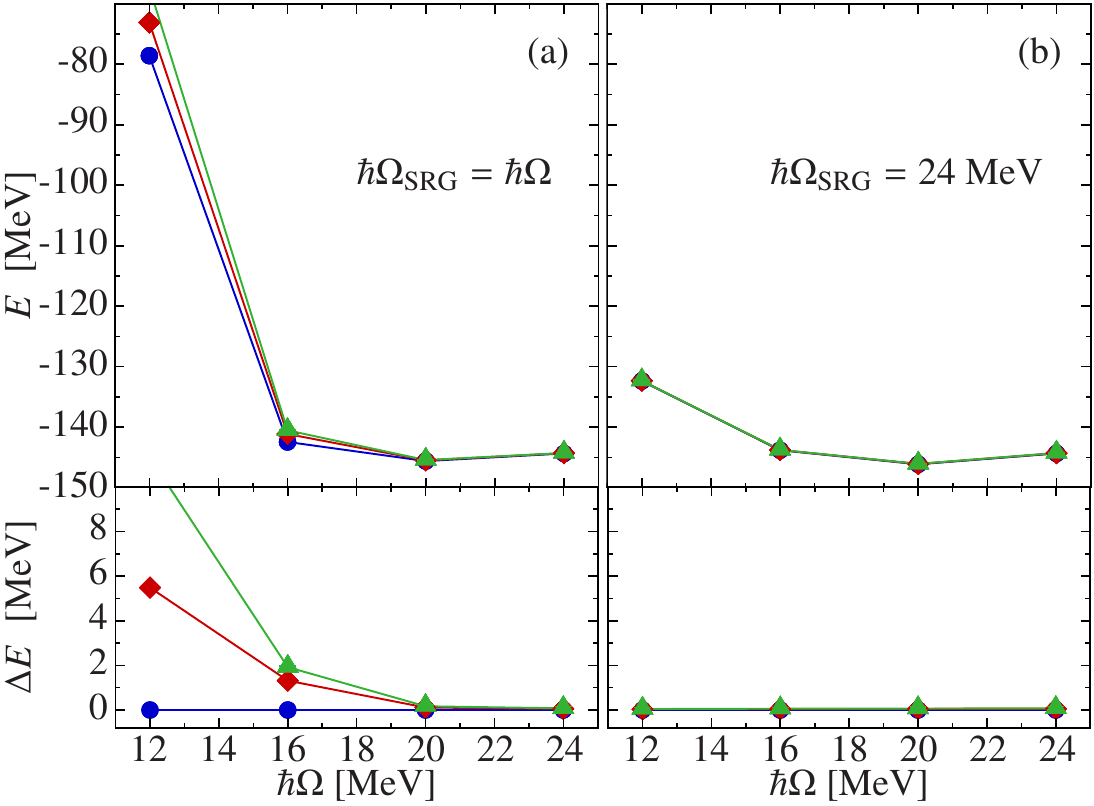}
\caption{(color online) Ground-state energy of \elem{O}{16} obtained at $N_{\max}=8$ for the NN+3N-full Hamiltonian with $\alpha=0.08\,\text{fm}^4$ as function of oscillator frequency $\hbar\Omega$. We compare the standard SRG evolution with $\hbar\Omega_{\text{SRG}}=\hbar\Omega$ (left column) with an SRG evolution at fixed $\hbar\Omega_{\text{SRG}}=24\,\text{MeV}$ and subsequent conversion of the matrix elements to the respective basis frequencies $\hbar\Omega$ (right column). The three curves correspond to the used SRG model space truncations defined by ramp A (\symbolcircle[FGBlue]), ramp B (\symboldiamond[FGRed]), and ramp C (\symboltriangle[FGGreen]). In the upper panels the absolute ground-state energies are plotted, while in the lower panels the deviations to energies obtained with ramp A are shown. 
}
\label{fig:prop_srgramp_conversion}
\end{figure}

In order to eliminate truncation artifacts at small basis frequencies $\hbar\Omega$ we use the frequency conversion introduced in Sec.~\ref{sec:srg_frequencyconv}. By using a larger frequency $\hbar\Omega_{\text{SRG}}$ for the SRG evolution and converting the evolved matrix elements afterwards to the nominal basis frequencies $\hbar\Omega$, we can remedy this problem completely. This is illustrated in Fig.~\ref{fig:prop_srgramp_conversion}, which shows the $\hbar\Omega$-dependence of the \elem{O}{16} ground-state energy at fixed $N_{\max}=8$ and $\alpha=0.08\,\text{fm}^4$ for the three different SRG model spaces. For the left-hand panels the three-body SRG-evolution is performed in an oscillator basis with the same $\hbar\Omega_{\text{SRG}}=\hbar\Omega$, for the right-hand panels we perform the SRG-evolution at fixed $\hbar\Omega_{\text{SRG}}=24\,\text{MeV}$ and convert to the basis frequency $\hbar\Omega$ of the many-body space subsequently. Note that the frequency conversion is performed using the same model-space truncation as for the solution of the SRG flow equations. The difference is obvious: Whereas a sizable dependence of the ground-state energy on the SRG ramp appears for the simple SRG evolution, the frequency-converted matrix elements do not show any dependence on the three-body model space, even when going to very low basis frequencies such as $\hbar\Omega=12\,\text{MeV}$. The direct comparison of the ground-state energies obtained without and with frequency conversion at the lowest frequency $\hbar\Omega=12\,\text{MeV}$ is particularly striking---the binding energy is dramatically underestimated by the SRG-transformed Hamiltonian without frequency conversion. Thus components of the initial Hamiltonians that are not captured by the three-body model space at $\hbar\Omega=\hbar\Omega_{\text{SRG}}=12\,\text{MeV}$ yield a large contribution to the binding energy. Without frequency conversion, calculations in this frequency domain, which is relevant, e.g., when trying to optimize the convergence of long-range operators, are not feasible.
 
With increasing mass number, the frequency range that is accessible without frequency conversion is reduced. Again we refer to our previous work in medium-mass nuclei, where this effect was already identified \cite{BiLa13,HeBo13}.

\subsection{Emergence of induced 4N interactions}

\begin{figure}[t]
\includegraphics[width=0.87\columnwidth]{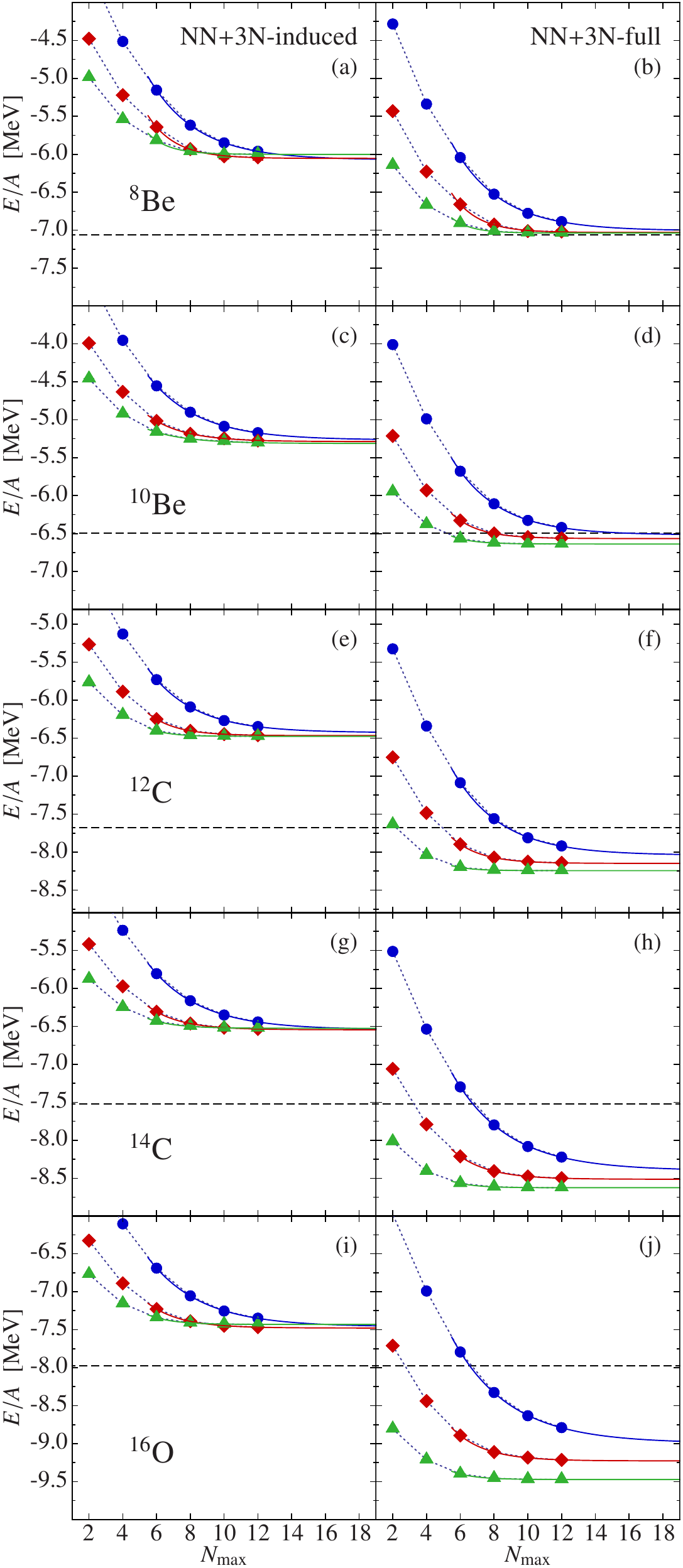}
\caption{(color online) Ground-state energies of \elem{Be}{8}, \elem{Be}{10}, \elem{C}{12}, \elem{C}{14}, and \elem{O}{16} (top to bottom) obtained with the NN+3N-induced (left column) and NN+3N-full Hamiltonian (right column) with $\alpha=0.04\,\text{fm}^4$ (\symbolcircle[FGBlue]), $0.08\,\text{fm}^4$ (\symboldiamond[FGRed]), and $0.16\,\text{fm}^4$ (\symboltriangle[FGGreen]) as function of $N_{\text{max}}$ for $\hbar\Omega=20\,\text{MeV}$. The dashed horizontal lines show experimental ground-state energies. }
\label{fig:alphaVar}
\end{figure}

After validating several technical aspects of the SRG evolution and the resulting Hamiltonians, we can now focus on one of the important side-effects of the SRG transformation---the emergence of induced many-body forces. The strong impact of SRG-induced 3N interactions when using an initial NN interactions was clearly demonstrated in Refs.~\cite{JuNa09,RoLa11,JuNa11,RoBi12} and many of the following calculations through the flow-parameter dependence of the NN-only results and the direct comparison with NN+3N-induced calculations. 

We have pointed out in Ref.~\cite{RoLa11} and reconfirmed this observation in Refs.~\cite{RoBi12, BiLa13} that beyond mid p-shell the calculations using the NN+3N-full Hamiltonian show a flow-parameter dependence of the ground-state energy, which is absent in  corresponding calculations with NN+3N-induced Hamiltonians. The systematic emergence of the flow-parameter dependence of the ground-state energy obtained with the NN+3N-full Hamiltonian is demonstrated in Fig.~\ref{fig:alphaVar} for isotopes in the mass range from $A=8$ to $16$. The left-hand column shows results for the NN+3N-induced Hamiltonian, the right-hand-column for the NN+3N-full Hamiltonian for three different flow parameters $\alpha=0.04$, $0.08$, and $0.16\,\text{fm}^4$ as function of the model-space truncation parameter $N_{\max}$. For all nuclei we are able to perform IT-NCSM calculations up to $N_{\max}=12$, which is sufficient to converge the ground-state energy for the softer Hamiltonians. We perform a simple exponential extrapolation of the energy using the last four data points to simplify the interpretation, the exponential fits are shown in Fig.~\ref{fig:alphaVar} as solid lines. 

Though the rate of convergence is different, the ground-state energies obtained with the NN+3N-induced Hamiltonians for different flow parameters all approach the same value in the limit $N_{\max}\to\infty$ to very good approximation. Thus, there is no indication that SRG-induced 4N terms, which formally exist, influence the ground-state energies---induced 4N contributions are negligible when starting from an initial chiral NN interaction.

The picture changes when including the initial chiral 3N interaction. For \elem{Be}{8} and lighter isotopes, the calculations with NN+3N-full Hamiltonians still do not exhibit a sizeable flow-parameter dependence of the converged ground-state energies. However, starting from mass $A\approx 10$ a flow-parameter dependence emerges, which increases systematically with $A$, both in absolute terms and in terms of the energy per nucleon. For \elem{O}{16}, the variation of the ground-state energy when going from $\alpha=0.04\,\text{fm}^4$ to $0.16\,\text{fm}^4$ reaches $0.5$ MeV per nucleon. It is driven by the initial 3N interaction, because the flow-parameter dependence is absent at the NN+3N-induced level. 

We stress that conclusions about the significance of induced many-body forces are valid only if the results are converged with respect to the relevant many-body truncations. For the IT-NCSM discussed here, this is just the model-space size $N_{\text{max}}$. For other methods this may be more complicated as we discussed previously in Refs.~\cite{BiLa13, HeBi13, HeBo13, RoBi12}. However, also these calculations confirm the aforementioned pattern for heavier nuclei. 

Keeping the influence of induced 4N interactions in mind, we can compare the ground-state energies to experiment, indicated by the dashed lines in Fig.~\ref{fig:alphaVar}. For the NN+3N-induced Hamiltonian, i.e., including initial chiral NN interactions only, we find an underbinding by $0.5$ to $1.2$ MeV per nucleon. This missing binding is provided by the chiral 3N interaction, i.e., at the level of the NN+3N-full Hamiltonian. For \elem{Be}{8} and \elem{Be}{10}, where induced 4N interactions are negligible, we find excellent agreement with the experimental binding energies. For \elem{C}{12}, \elem{C}{14}, and \elem{O}{16} the NN+3N-full calculations show an increasing flow-parameter dependence and an increasing overbinding. Although a sizable part of the overbinding seems to be due to the missing SRG-induced 4N contributions, based on these calculations, we cannot decide whether all of the overbinding is of this origin or whether it is resulting from deficiencies of the initial Hamiltonian.

We conclude that starting from mid-p-shell, SRG-induced 4N interactions (or even higher-order contributions) start to have an impact on ground-state energies as soon as we include the standard chiral 3N interaction in the initial Hamiltonian. At this moment we have to discard these induced higher-order many-body forces, but efforts to account for SRG-induced 4N interactions are currently under way. Excitation energies, however, do not show a sizable flow-parameter dependence once convergence with respect to $N_{\max}$ is reached, as shown in Refs.~\cite{RoLa11,MaAk13,MaVa13x}.

\subsection{Origin of the induced 4N interactions}
\label{sec:origin}

Having identified the initial chiral 3N interactions as the origin of sizable SRG-induced 4N contributions, we further analyze the role of the individual parts of the N$^2$LO 3N interaction. The 3N interaction is usually split into a two-pion exchange, a two-nucleon contact one-pion exchange and a three-nucleon contact term. The corresponding operator structures are 
\eq{\label{eq:tpe}
\sum_{i\neq j\neq k} \sum_{\alpha,\beta} \frac{1}{2} \bigg( \frac{g_A}{2 F_\pi} \bigg)^2 
\frac{(\vec{\sigma}_i \cdot \vec{q}_i)(\vec{\sigma}_j \cdot \vec{q}_j)}{(\vec{q}_i^2 + M_\pi^2)(\vec{q}_j^2 + M_\pi^2)} F_{ijk}^{\alpha\beta} \tau_{i}^{\alpha} \tau_{j}^{\beta} 
}
with
\eq{
F_{ijk}^{\alpha\beta} = \delta^{\alpha\beta} 
\left[ -\frac{4c_1 M_\pi^2}{F_\pi^2} +\frac{2 c_3 }{F_\pi^2} \vec{q}_i\cdot\vec{q}_j \right] 
+   \displaystyle{\sum_{\gamma}} \frac{c_4}{F_\pi^2} \epsilon^{\alpha\beta\gamma} \tau_k^{\gamma} \vec{\sigma}_k\cdot [\vec{q}_i\times\vec{q}_j] 
}
for the two-pion exchange term depending on the low-energy constants $c_1$, $c_3$, and $c_4$ (or $c_i$ for short),
\eq{\label{eq:ope}
  - c_D \sum_{i\neq j\neq k} \frac{g_A}{8F_\pi^{4}\Lambda_{\chi}}  \frac{\vec{\sigma}_j\cdot\vec{q}_{j}}{\vec{q}  _j^{2} +M_\pi^{2}} (\vec{\tau}_i\cdot\vec{\tau}_j)(\vec{\sigma}_i\cdot\vec{q}_j) 
}
for the two-nucleon contact one-pion exchange term proportional to low-energy constant $c_D$, and
\eq{\label{eq:3nc}
  c_E\sum_{j\neq k} \frac{1}{2 F_\pi^4 \Lambda_\chi} (\vec{\tau}_j\cdot\vec{\tau}_k)
}
for the three-nucleon contact term with strength $c_E$. Here we adopt the notation and constants of Ref.~\cite{EpNo02}.
In order to assess the impact of the various terms on the SRG-induced 4N interactions we switch off the terms individually by setting the respective low-energy constant to zero. For each case, we refit $c_E$ to reproduce the \elem{He}{4} ground-state energy of $-28.30$ MeV with an uncertainty below $10$ keV in NCSM calculations with the bare Hamiltonian. We keep $c_D=-0.2$ as determined from the triton $\beta$-decay half-life, except for the case with $c_E=0$ where $c_D$ is used to fit the \elem{He}{4} energy. The different sets of low-energy constants obtained from the fit are summarized in Tab.~\ref{tab:LECs}. The resulting Hamiltonians, which are still fixed entirely in the three- and four-body system, are evolved consistently in the SRG framework and enter into the IT-NCSM calculations.

\begin{table}[b]
\caption{Low-energy constants of the chiral 3N interaction at N${}^2$LO for the standard interaction \cite{GaQu09} and different variants described in the text. All variants are refit in NCSM calculations with the bare interactions to reproduce the experimental \elem{He}{4} ground-state energy. }
\label{tab:LECs}
\begin{ruledtabular}
\begin{tabular}{c|c c c c c c}
       &  $\Lambda_{\text{3N}}$  &  $c_1$ & $c_3$  &  $c_4$  & $c_D$  & $c_E$ 	\\
       &  [MeV/$c$]       &  [$\text{GeV}^{-1}$]  &  [$\text{GeV}^{-1}$]  &  [$\text{GeV}^{-1}$]  &   &  \\
\hline
standard & 500  & -0.81 & -3.2   &  5.4    &    -0.2  &    -0.205   \\ 
\hline
$c_i=0$  & 500  &  0        & 0        & 0        &    -0.2       &    0.444      \\
$c_D=0$  & 500  & -0.81 & -3.2   &  5.4       &       0   &   -0.205       \\
$c_E=0$  & 500  & -0.81 & -3.2   &  5.4       &       1.238  &   0         \\
\hline
$c_1=0$  & 500  &  0 & -3.2   &  5.4       &       -0.2  &   -0.207        \\
$c_3=0$  & 500  & -0.81 & 0   &  5.4       &       -0.2  &   -0.228      \\
$c_4=0$  & 500  & -0.81 & -3.2   &  0       &       -0.2  &  0.141       \\
\hline
$\Lambda_{\text{3N}}=450$ &  450   & -0.81 & -3.2   &  5.4    &    -0.2  &   -0.016   \\
$\Lambda_{\text{3N}}=400$ &  400   & -0.81 & -3.2   &  5.4    &    -0.2  &   0.098  \\
$\Lambda_{\text{3N}}=350$ &  350   & -0.81 & -3.2   &  5.4    &    -0.2  &   0.205  \\
\end{tabular}
\end{ruledtabular}
\end{table}

\begin{figure}[t]
\includegraphics[width=\columnwidth]{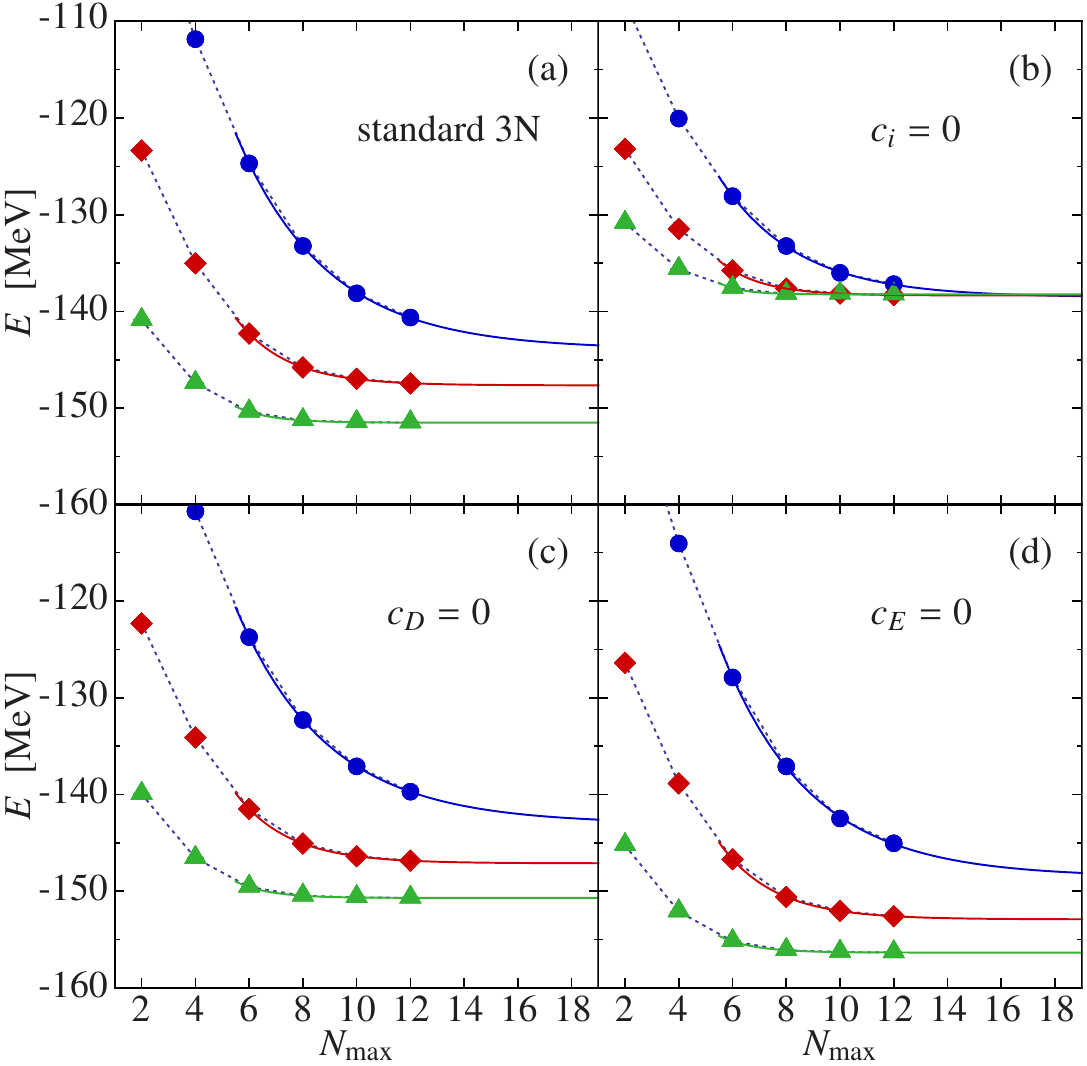}
\caption{(color online) Ground-state energy of \elem{O}{16} obtained with the NN+3N-full Hamiltonian with $\alpha=0.04\,\text{fm}^4$ (\symbolcircle[FGBlue]), $0.08\,\text{fm}^4$ (\symboldiamond[FGRed]), and $0.16\,\text{fm}^4$ (\symboltriangle[FGGreen]) as function of $N_{\text{max}}$. Results for the standard Hamiltonian are shown in panel (a), and those for $c_i=0$, $c_D=0$, and $c_E=0$ in panels (b), (c), and (d) respectively.}\label{fig:O16cicDcEanalysis}
\end{figure}
\begin{figure}[t]
\includegraphics[width=\columnwidth]{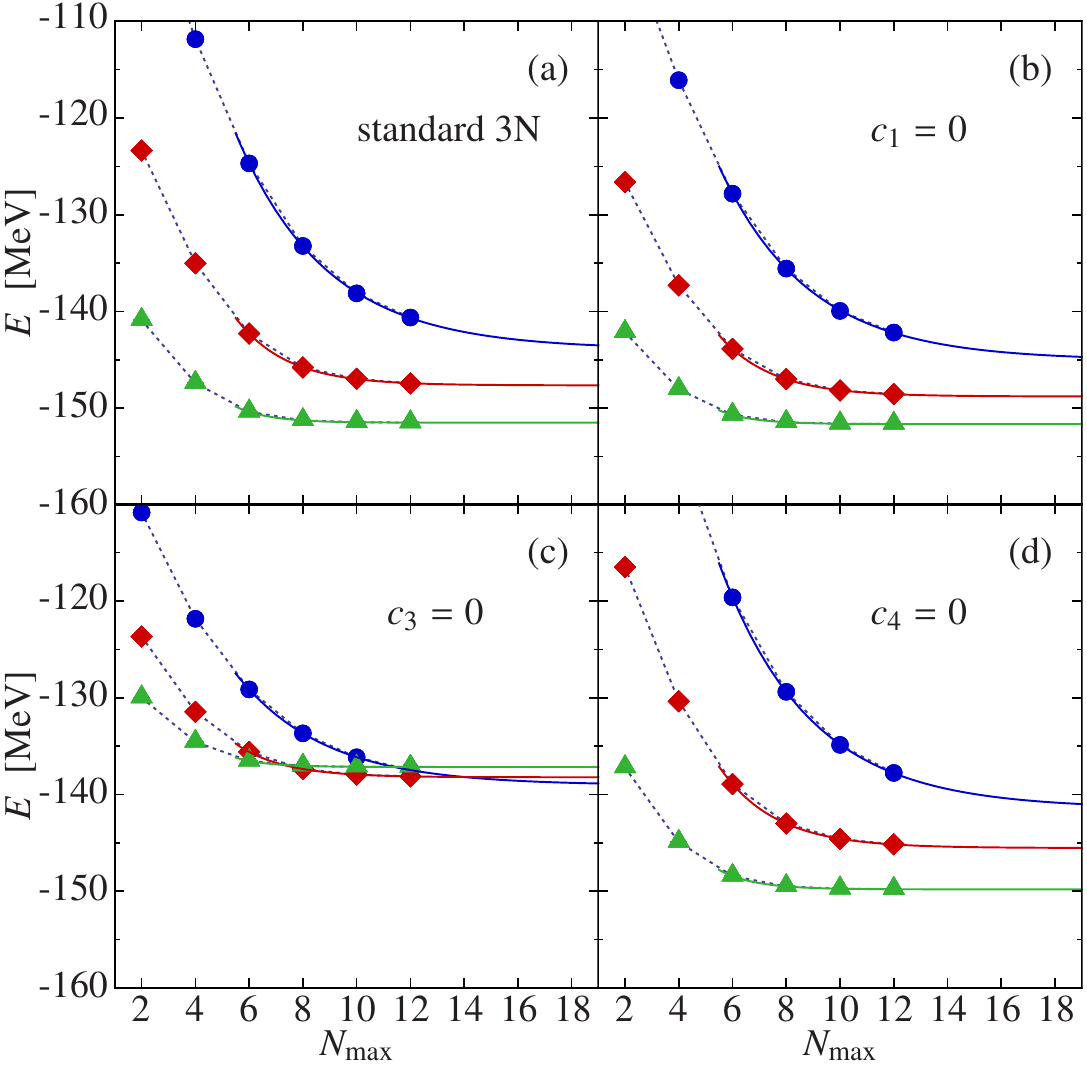}
\caption{(color online) Ground-state energy of \elem{O}{16} obtained with the NN+3N-full Hamiltonian with $\alpha=0.04\,\text{fm}^4$ (\symbolcircle[FGBlue]), $0.08\,\text{fm}^4$ (\symboldiamond[FGRed]), and  $0.16\,\text{fm}^4$ (\symboltriangle[FGGreen]) as function of $N_{\text{max}}$. Results for the standard Hamiltonian are shown in panel (a), and those for $c_1=0$, $c_3=0$, and $c_4=0$ in panels (b), (c), and (d), respectively.}
\label{fig:O16c1c3c4analysis}
\end{figure}

We apply these modified 3N interactions in a series of ground-state calculations for \elem{O}{16} up to $N_{\max}=12$ with the three flow parameters $\alpha=0.04$, $0.08$ and $0.16\,\text{fm}^{4}$. The results for the modified Hamiltonians with $c_i=0$, $c_D=0$, and $c_E=0$ are summarized in Fig.~\ref{fig:O16cicDcEanalysis}. In panel (a) the ground-state energies obtained with the standard Hamiltonian, showing the flow-parameter dependence discussed in the previous section, are depicted for comparison. When switching off the two-nucleon contact one-pion exchange contribution ($c_D=0$) or the three-nucleon contact term ($c_E=0$) there is no sizable change of the flow-parameter dependence as compared to the standard Hamiltonian, as seen in Fig.~\ref{fig:O16cicDcEanalysis}(c) and (d), respectively. Thus, neither of these two terms of the chiral 3N interaction drives the SRG-induced many-body forces. The picture changes dramatically, if we switch off the two-pion exchange terms ($c_i=0$). As depicted in Fig.~\ref{fig:O16cicDcEanalysis}(b), the flow-parameter dependence of the converged ground-state energy vanishes completely in this case. Thus, the long-range two-pion terms in the chiral 3N interaction alone are responsible for the emergence of sizable induced many-body contributions throughout the SRG evolution.

We can carry this analysis even further and investigate the role of the three different two-pion exchange contributions by switching-off the $c_1$, $c_3$, and $c_4$ terms individually. The resulting ground-state energies for \elem{O}{16} are depicted in Fig.~\ref{fig:O16c1c3c4analysis}. The comparison with the flow-parameter dependence of the standard Hamiltonian shows that the $c_1$ contribution does not affect the induced many-body terms. Also, switching off the $c_4$ term only causes minor changes in the flow-parameter dependence. However, eliminating the $c_3$ of the chiral 3N interaction leads to a drastic reduction of the flow-parameter dependence, as shown in Fig.~\ref{fig:O16c1c3c4analysis}(c). We conclude that the $c_3$ contribution is the major driver for the induced beyond-3N terms in the SRG evolution. 

Because of their complicated operator structure, including intermediate-range tensor- and spin-orbit-type interactions, the $c_i$ terms are likely candidates for causing many-body correlations that give rise to induced many-body interactions in the SRG evolution---in analogy to the tensor interaction at the NN-level as an important source of induced 3N contributions \cite{RoNe10}. However, it is not obvious why the $c_3$ contribution is the dominant source and the $c_4$ term contributes so little. In contrast to the $c_1$ term, which contributes very little to the ground-state energies of \elem{He}{4} or \elem{O}{16}, the contribution of the $c_4$ term to the ground-state energy is not small. This can be seen from the large change of $c_E$ that is necessary to reproduce the \elem{He}{4} ground-state energy when $c_4$ is set to zero. 

These findings might prove useful for the design of alternative SRG generators which aim to suppress the induced many-body terms. However, initial attempts along these lines were not successful.

\subsection{Reduced initial three-nucleon cutoff}

\begin{figure*}[t]
\includegraphics[width=0.9\textwidth]{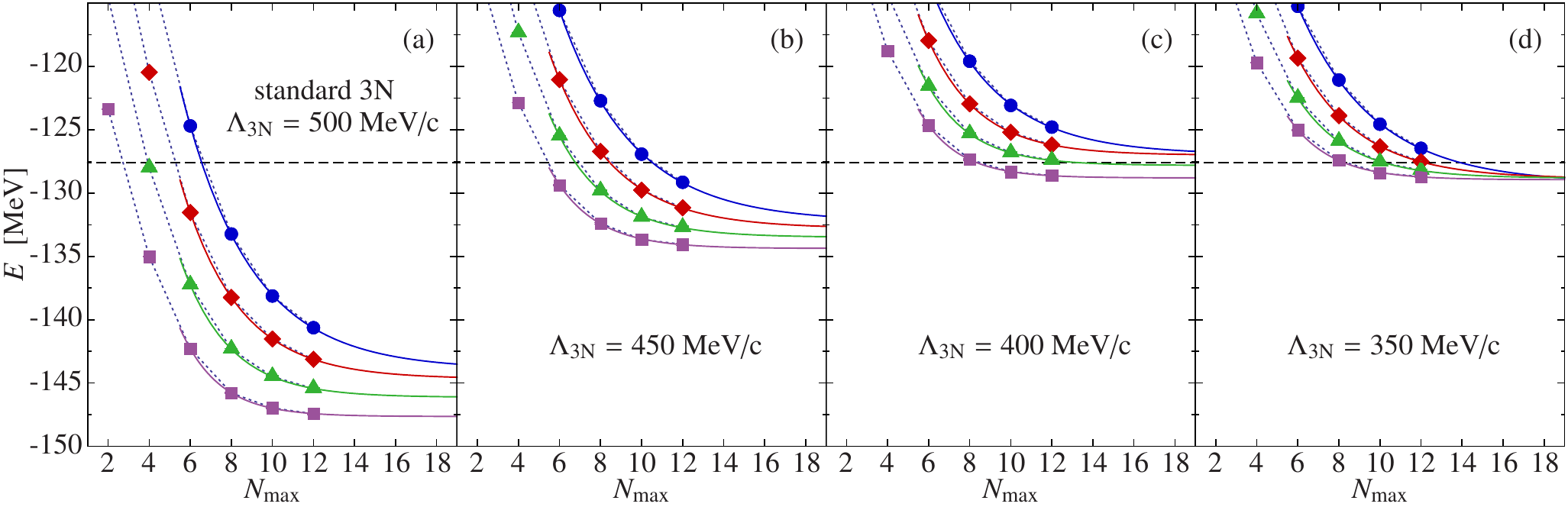}
\caption{(color online) Dependence on the cutoff of the 3N interaction $\Lambda_{\text{3N}}$ of the \elem{O}{16} ground-state energy obtained with the NN+3N-full Hamiltonian with $\alpha=0.04\,\text{fm}^4$ (\symbolcircle[FGBlue]), $0.05\,\text{fm}^4$ (\symboldiamond[FGRed]), $0.0625\,\text{fm}^4$ (\symboltriangle[FGGreen]), and $0.08\,\text{fm}^4$ (\symbolbox[FGViolet]). Results for the standard Hamiltonian are shown in panel (a), and those for $\Lambda_{\text{3N}}=450\,\text{MeV/c}$, $400\,\text{MeV/c}$, and $350\,\text{MeV/c}$ in panels (b), (c) and (d), respectively.}
\label{fig:Lambda-dependence}
\end{figure*}

Motivated by the observation that small modifications of the structure of the initial chiral 3N interaction can eliminate the SRG-induced many-body interactions, we study the behavior of the flow-parameter dependence of the \elem{O}{16} ground-state energy as function of the three-body cutoff $\Lambda_{\text{3N}}$ used for the regularization of the chiral 3N interaction at N$^2$LO. As outlined in the previous section, we refit the $c_E$ parameter for each initial 3N cutoff to reproduce the \elem{He}{4} ground state energy in NCSM calculations with the bare Hamiltonian. The resulting values of $c_E$ for cutoffs in the range from $\Lambda_{\text{3N}}=350$ to $500\,\text{MeV/c}$ are summarized in Tab.~\ref{tab:LECs}.
  
The IT-NCSM results for the ground-state energies of \elem{O}{16} are presented in Fig.~\ref{fig:Lambda-dependence} for the different initial 3N cutoffs. The flow-parameter dependence of the converged energies shows a clear systematics: with decreasing cutoff $\Lambda_{\text{3N}}$ the flow-parameter dependence is rapidly reduced. For $\Lambda_{\text{3N}}=350\,\text{MeV/c}$ the converged ground-state energies exhibit no flow-parameter dependence in the range from $\alpha=0.04$ to $0.08\,\text{fm}^4$ anymore. Already at $\Lambda_{\text{3N}}=400\,\text{MeV/c}$ the ground-state energies only vary by about 2\% over this flow-parameter range. In combination with the analysis of Sec.~\ref{sec:origin}, we can conclude that the higher-momentum components, i.e., contributions that are eliminated by lowering the 3N cutoff to $\Lambda_{\text{3N}}=350\,\text{MeV/c}$, of the two-pion terms of the 3N interaction are responsible for the emergence of SRG-induced 4N interactions.

As the flow-parameter dependence decreases, the \elem{O}{16} ground-state energy systematically approaches the experimental binding energy. For both, $\Lambda_{\text{3N}}=350$ and $400\,\text{MeV/c}$ the calculated energies agree very well with experiment. This is remarkable, since no experimental data beyond $A=4$ was used to constrain these Hamiltonians. Since the flow-parameter dependence and thus the contribution of induced beyond-3N interactions is small, we can conclude that these reduced-cutoff Hamiltonians enable a parameter-free description of the \elem{O}{16} ground-state energy. This finding is confirmed in a systematic study of the ground states of even oxygen isotopes from \elem{O}{12} to \elem{O}{26} using the IT-NCSM, coupled-cluster theory, and the newly developed multi-reference in-medium SRG \cite{HeBi13}. We have shown that the chiral 3N interactions with reduced cutoff can well reproduce the experimental ground-state energies throughout the oxygen isotopic chain and describe the position of the dripline correctly without any phenomenological adjustments. Furthermore, for medium-mass nuclei, like calcium and nickel isotopes, the coupled-cluster calculations discussed in Refs.~\cite{RoBi12,BiLa13} indicate that these interactions still provide a remarkably good description of ground-state energies.
  
Of course, lowering the cutoff too far will eliminate physically important components of the interaction. First indications are already seen for the interaction with $\Lambda_{\text{3N}}=400\,\text{MeV/c}$ in the spectroscopy of p-shell nuclei for observables that depend sensitively on the 3N interaction. A prime example is the ordering of the lowest states in \elem{B}{10}: the standard chiral 3N interaction with $\Lambda_{\text{3N}}=500\,\text{MeV/c}$ predicts the ground-state to be a $3^+$ with an approximately correct excitation energy to the first $1^+$ state. Reducing the 3N cutoff to $\Lambda_{\text{3N}}=400\,\text{MeV/c}$ gives almost degenerate $3^+$ and $1^+$ states with a tendency for the $1^+$ to become the ground state. However, one should note that also the standard 3N interaction has deficiencies regarding p-shell spectroscopy. The excitation energy of the first $1^+$ state in \elem{C}{12} is underestimated by about 4 MeV for $\Lambda_{\text{3N}}=500\,\text{MeV/c}$, but is within 0.5 MeV of the experimental value for $\Lambda_{\text{3N}}=400\,\text{MeV/c}$.
These and related effects of the 3N interaction on the spectroscopy of p-shell nuclei will be discussed in forthcoming publications \cite{MaVa13x}.

\section{Comparison and Extrapolation}
\label{sec:comp}

We close this discussion with a comparison of our results for ground-state energies of p-shell nuclei with a set of similar calculations by Jurgenson et al. \cite{JuMa13}. These authors are using the same standard chiral NN+3N Hamiltonian as starting point and they also use the SRG evolution and the NCSM to tackle the many-body problem. However, there are significant differences regarding (i) the model space for the SRG evolution, (ii) the handling of the 3N matrix elements, and (iii) the solution of the many-body problem: 

\begin{enumerate}

\item[(i)] We employ a different truncation pattern for the three-body Jacobi-HO model-space of the SRG evolution as discussed in Sec.~\ref{subsec:role_srg_space}, allowing for larger spaces for the $J=3/2$ and $5/2$ partial waves as compared to Jurgenson et al. More importantly, we use the frequency conversion discussed in Sec.~\ref{sec:srg_frequencyconv}, i.e., the SRG evolution is performed for fixed frequency $\hbar\Omega_{\text{SRG}}=24$ MeV and we convert the resulting matrix elements to all other basis frequencies of interest. This eliminates the truncation artifacts at low frequencies, as demonstrated in Sec. \ref{subsec:role_srg_space}.

\item[(ii)] We use the $JT$-coupled scheme for handling the 3N matrix elements instead of the $m$-scheme storage used by Jurgenson et al. This enables us to precompute and store 3N matrix-element sets for much larger spaces, as highlighted in Sec. \ref{subsec:computational_strategy}. For an $N_{\max}=8$ calculation of \elem{C}{12}, corresponding to $E_{3\max}=11$, the $m$-scheme approach requires about 33 GB for the 3N matrix elements in single precision \cite{MaAk13}. In the $JT$-coupled approach we need only 0.4 GB with about the same performance for retrieving individual $m$-scheme three-body matrix elements, because of our highly efficient decoupling algorithm. We can routinely generate $JT$-coupled matrix element sets up to $E_{3\max}=16$, which is sufficient for $N_{\max}=13$ calculations in \elem{C}{12} and requires only 10 GB of storage. 

\item[(iii)]
We use the importance truncation to extend the reach of the NCSM. The limit of full NCSM calculations with NN+3N Hamiltonians for \elem{C}{12} today is at $N_{\max}=8$ or $9$ (see \cite{MaVa13x}). With the IT-NCSM we can easily extend the ground-state calculations up to $N_{\max}=12$ at a fraction of the computational cost of full NCSM calculations at $N_{\max}=8$. In combination with SRG-evolved Hamiltonians, the gain from $N_{\max}=8$ to $N_{\max}=12$ is important, since it brings us sufficiently close to convergence so that different possible extrapolation schemes become more robust and accurate. 

\end{enumerate}

\begin{figure}
\includegraphics[width=0.75\columnwidth]{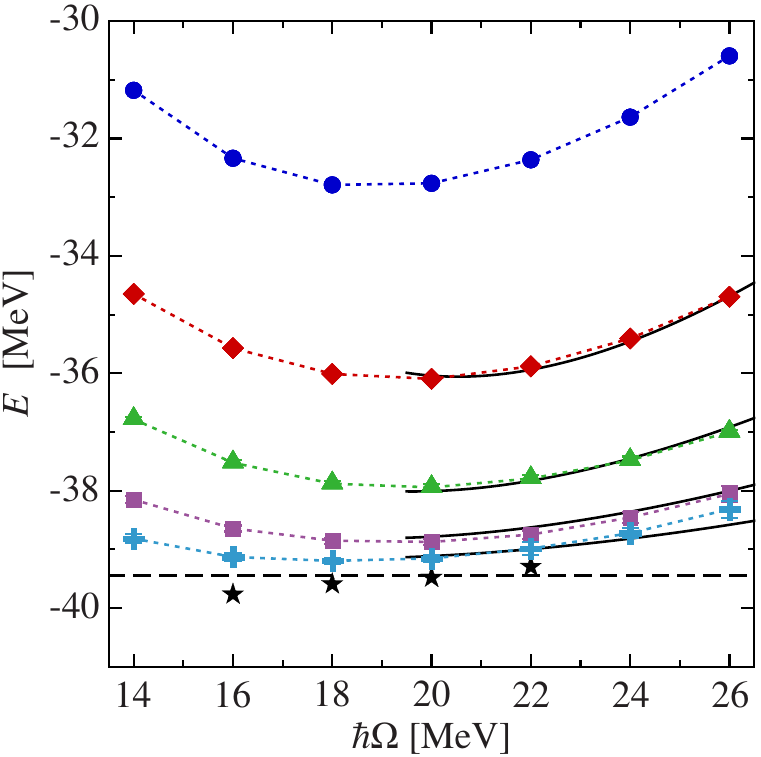}
\caption{(color online) Ground-state energy of \elem{Li}{7} as function of basis frequency $\hbar\Omega$ obtained with NN+3N-full Hamiltonian with $\alpha=0.0625\,\text{fm}^4$ and $\Lambda_{\text{3N}}=500\,\text{MeV}/c$. We use frequency-converted 3N matrix-elements with an SRG evolution performed at $\hbar\Omega_{\text{SRG}}=24\,\text{MeV}$. The different symbols correspond to $N_{\max}=4$ (\symbolcircle[FGBlue]), $6$ (\symboldiamond[FGRed]), $8$ (\symboltriangle[FGGreen]), $10$ (\symbolbox[FGViolet]), and $12$ (\symbolcross[FGLightBlue]) with error bars extracted from the threshold extrapolation. The solid lines show the IR-UV fit using the results in the window from $\hbar\Omega=20$ to $26$ MeV, the dashed horizontal line shows the $N_{\max}\to\infty$ ground-state energy resulting from this fit. The black stars show the results of simple extrapolations at fixed $\hbar\Omega$ (see text).}
\label{fig:itncsm_Li7_GS}
\end{figure}
\begin{figure}
\includegraphics[width=0.75\columnwidth]{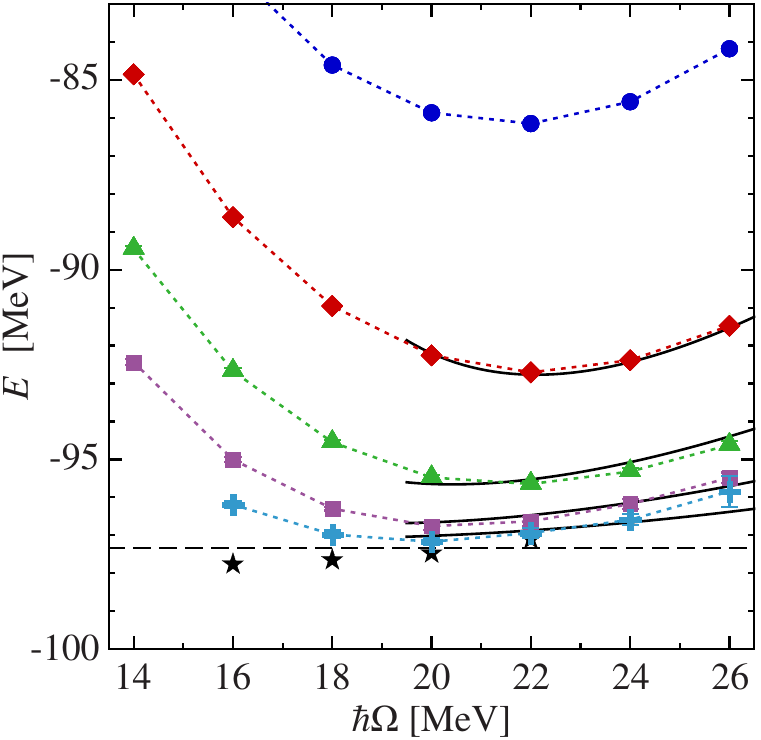}
\caption{(color online) Same as Fig.~\ref{fig:itncsm_Li7_GS} for the ground-state energy of \elem{C}{12}.}
\label{fig:itncsm_C12_GS}
\end{figure}
\begin{figure}
\includegraphics[width=0.75\columnwidth]{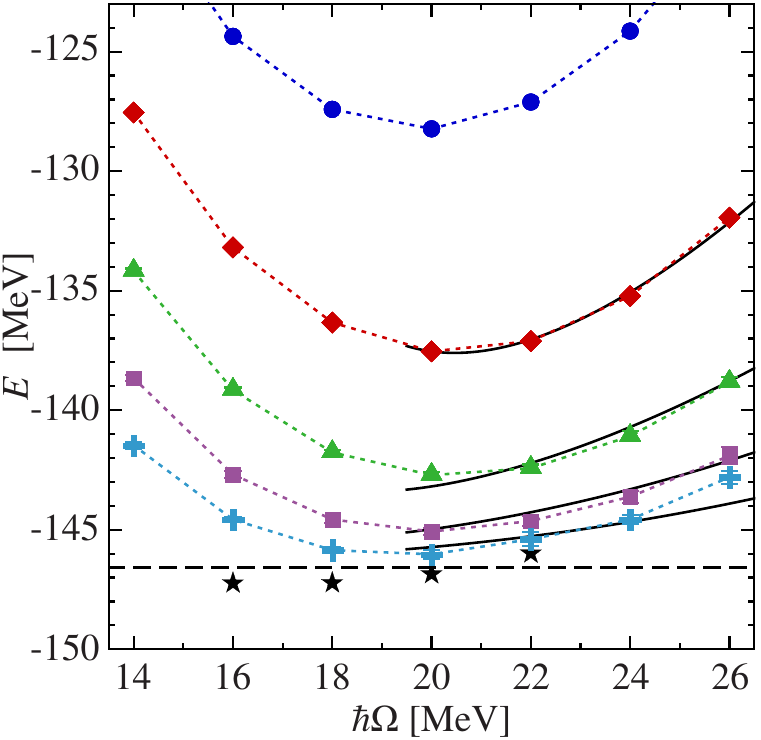}
\caption{(color online) Same as Fig.~\ref{fig:itncsm_Li7_GS} for the ground-state energy of \elem{O}{16}.}
\label{fig:itncsm_O16_GS}
\end{figure}

Two examples for ground-state calculations that can be compared directly to the work of Jurgenson et al. are presented in Figs.~\ref{fig:itncsm_Li7_GS} and \ref{fig:itncsm_C12_GS}. In Fig.~\ref{fig:itncsm_Li7_GS} we show the convergence of  of the ground-state energy of \elem{Li}{7} with increasing $N_{\max}=4,6,...,12$ as function of the basis frequency $\hbar\Omega$ obtained with the NN+3N-full Hamiltonian for $\alpha=0.0625\,\text{fm}^4$ --- corresponding to Fig. 15 of Ref.~\cite{JuMa13}. We emphasize that because of the frequency conversion, also the results at low $\hbar\Omega$ are accurate. It is evident that the $N_{\max}=12$ results are already very close to convergence and provide an excellent starting point for robust and accurate extrapolations. The corresponding ground-state energies for \elem{C}{12} are presented in Fig.~\ref{fig:itncsm_C12_GS} and can be compared to Fig. 16 of Ref.~\cite{JuMa13}. Even for this mid p-shell nucleus we can perform the IT-NCSM calculations up to $N_{\max}=12$ which is already very close to the converged result. For completeness, we show in Fig.~\ref{fig:itncsm_O16_GS} the results for \elem{O}{16} ground-state energies with the same Hamiltonian, which have not been discussed in Ref.~\cite{JuMa13}, again reaching up to $N_{\max}=12$ and thus close to convergence. 

Even the simplest extrapolation scheme, using the three-parameter exponential ansatz $E(N_{\max}) = E_{\infty} + a \exp(-b N_{\max})$ and fitting to three or four large-$N_{\max}$ results at a single frequency $\hbar\Omega$, provides robust results. In Tab.~\ref{tab:extrapolation} we summarize the extrapolated energies $E_{\infty}$ for various frequencies. Here we use the four largest $N_{\max}$ results for the fit in order to stabilize the extrapolation against uncertainties resulting from the threshold extrapolation of the individual IT-NCSM calculations for the different $N_{\max}$. There is a slight systematic dependence of the results on the basis frequency in all cases, tending to reduce the binding energy with increasing $\hbar\Omega$. However, comparing the results at the optimal frequency, which provides the minimum energy in the largest model space, with the neighboring frequencies we observe differences below 0.5\%. Keeping in mind the uncertainties that result from the importance truncation and threshold extrapolation, which are of similar magnitude, we consider this simple extrapolation at fixed optimal frequency as sufficiently accurate once the largest $N_{\max}$ is close to convergence. The IT-NCSM is instrumental to reach these large $N_{\max}$ values and we can limit ourselves to the simple extrapolation scheme. 

More sophisticated and theoretically better motivated extrapolation schemes were proposed in several recent works \cite{CoAv12,FuHa12,MoEk13}. They take the high-momentum (UV) and long-range (IR) truncations implied by a finite HO basis into account for the construction of an extrapolation function in a framework inspired by effective field theory. Though the quantitative exploration of these extrapolation schemes is only beginning, we employ the IR-UV extrapolation scheme for the energy in the formulation proposed in Ref.~\cite{MoEk13} for comparison. We use $E(N_{\max},\hbar\Omega) = E_{\infty} + a_1 \exp(-2 b_1 \Lambda_{\text{UV}}) + a_2 \exp(-2 b_2 L_2)$, with $\Lambda_{\text{UV}} = \sqrt{2 (e_{\max}+3/2)}/a_{\text{HO}}$ and $L_2=a_{\text{HO}} \sqrt{2 (e_{\max}+3/2+2)}$, where $a_{\text{HO}}$ is the oscillator length and $e_{\max}$ the maximum single-particle energy quantum number represented in the basis, i.e., $e_{\max}=N_{\max}+1$ for p-shell nuclei. We note that all points of the selected subset enter our fits with equal weight, while alternative extrapolation methods \cite{MaVa09} have employed increased weights for data points closer to the converged results.

The results of the IR-UV extrapolation summarized in Tab.~\ref{tab:extrapolation}. Again we have to select a range in $N_{\max}$ and $\hbar\Omega$ for the data entering into the fit. As for the simple exponential extrapolation  we use the four largest $N_{\max}$ results and a range of frequencies up to the maximum available frequency of $\hbar\Omega=26$ MeV. Since the theoretical foundation of the extrapolation scheme is more solid in the UV regime, i.e., towards the high-frequency side of the energy minimum, we vary the low-frequency end of the data set included in the fit around the minimum to probe the stability of the extrapolation.  

Based on the same $N_{\max}$-range as input data, the IR-UV extrapolation also exhibits as systematic dependence on the frequency-range included in the fit. As expected, the dependence is somewhat smaller than for the simple extrapolations at a single frequency. The comparison of the simple extrapolation at the optimal frequency, i.e., $\hbar\Omega=18$ MeV for \elem{Li}{7} and $\hbar\Omega=20$ MeV for \elem{C}{12} and \elem{O}{16}, with the IR-UV extrapolation based on the high-frequency data $\hbar\Omega=20 - 26$ MeV, reveals nice agreement. One should note, however, that the IR-UV extrapolation for the heavier nuclei does not fully capture the curvature of the energy as function of $\hbar\Omega$ at fixed $N_{\max}$, as can be seen from the comparison of data and fit function in Figs.~\ref{fig:itncsm_C12_GS} and \ref{fig:itncsm_O16_GS}. These deviations are getting worse as more data points at lower $\hbar\Omega$ are included. Further investigations into the these extrapolation methods in the upper p-shell are certainly desirable.  

\begin{table}
\caption{Extrapolated ground-state energies $E_{\infty}$ in [MeV] of \elem{Li}{7}, \elem{C}{12}, and \elem{O}{16} using the NN+3N-full Hamiltonian at $\alpha=0.0625\,\text{fm}^4$ for different extrapolation schemes and subsets of the IT-NCSM results presented in Figs.~\ref{fig:itncsm_Li7_GS}, \ref{fig:itncsm_C12_GS}, and \ref{fig:itncsm_O16_GS} (see text). 
}
\label{tab:extrapolation}
\begin{ruledtabular}
\begin{tabular}{c c c c c c}
 & $N_{\max}$ & $\hbar\Omega$ [MeV] & \elem{Li}{7} & \elem{C}{12} & \elem{O}{16} \\
\hline
simple & $6-12$ & $16$     &  -39.77  &  -97.76  & -147.23 \\
&$6-12$ & $18$      &  -39.59  &  -97.64  & -147.22 \\
&$6-12$ & $20$      &  -39.48  &  -97.47  & -146.85 \\
&$6-12$ & $22$      &  -39.30  &  -97.10  & -145.98 \\
\hline
IR-UV & $6-12$ & $14-26$    &  -39.66  &  -97.04  & -145.44 \\
&$6-12$ & $16-26$    &  -39.61  &  -97.10  & -145.78 \\
&$6-12$ & $18-26$    &  -39.54  &  -97.26  & -146.26 \\
&$6-12$ & $20-26$    &  -39.45  &  -97.33  & -146.59 \\
\hline
IR-UV &$2-8$ & $14-26$    &  -39.43  &  -97.28  & -144.23 \\
&$2-8$ & $16-26$    &  -40.19  &  -98.79  & -148.61 \\
&$2-8$ & $18-26$    &  -40.72  &  -99.92  & -152.88 \\
&$2-8$ & $20-26$    &  -40.98  & -100.43  & -158.13 \\
\end{tabular}
\end{ruledtabular}
\end{table}

Even for the IR-UV extrapolation, the availability of input data close to convergence is important. If we ignore the results for $N_{\max}=10$ and $12$ and repeat the analysis using the range from $N_{\max}=2-8$ as input, the sensitivity of the extrapolated energies on the choice of the frequency-range increases by an order of magnitude as shown in the lower part of Tab.~\ref{tab:extrapolation}. Thus, even with improved extrapolation tools the additional steps in $N_{\max}$ that the IT-NCSM offers are vital to obtain robust results within our fitting strategy. 

The IR-UV extrapolation scheme using preferentially large frequencies entails a significant increase in computational cost, since the dimension of the importance-truncated model space grows with increasing basis frequency, as many more basis states with small amplitudes need to be superimposed to build-up the net size of the nucleus. This makes the calculations for individual importance thresholds $\kappa_{\min}$ more demanding and increases the uncertainties of the threshold extrapolations. Since the IT-NCSM allows us to reach sufficiently large $N_{\max}$, we typically use the simple extrapolation at and around the optimal frequency in practical applications.

\section{Conclusions}

We have discussed a chain of developments enabling ab initio nuclear structure calculations for light and medium-mass nuclei using SRG-evolved chiral NN+3N Hamiltonians in large many-body model spaces. By introducing a new $JT$-coupled storage scheme for the 3N matrix elements together with a fast on-the-fly decoupling in the many-body calculation, we are able to reach model spaces of unprecedented size with explicit 3N interactions. It turns out that controlling the truncation uncertainties of the SRG-evolved Hamiltonians is one of the most critical elements for ab initio calculations beyond the lightest isotopes. 

A first truncation uncertainty results from the finite Jacobi-HO model space used to perform the SRG-evolution of the 3N interaction. The effect of this truncation is amplified with increasing mass number and affects low basis frequencies in particular. We introduced a simple frequency conversion of the 3N matrix elements to fix this issue for nuclei in the p- and sd-shell. However, one has to revisit the role of this truncation when going to medium-mass and heavy nuclei. A second truncation uncertainty results from the omission of SRG-induced four- and multi-nucleon interactions, which become significant beyond mid p-shell. Apart from the explicit inclusion of SRG-induced 4N interactions, which is under investigation at the moment, one can remedy this issue by using chiral interactions with lower initial cutoffs. It would be very beneficial for applications of next generation chiral Hamiltonians, if a sequence of cutoffs extending as low as $400$ MeV/c would be available. Various attempts to design alternative SRG-generators that suppress induced 4N terms but retain the favorable convergence behavior of the standard generator have not been successful so far. 

When going beyond NCSM-type calculations, additional truncations of the Hamiltonian have to be introduced. Present medium-mass approaches, e.g., coupled-cluster theory, typically work in model spaces obtained from a finite set of Hartree-Fock single-particle states, which are not compatible with the $E_{3\max}$ truncation of the 3N matrix elements. Furthermore, truncations of the normal-ordered Hamiltonian at the two-body level are being used to avoid the generalization of the formalism to explicit 3N contributions. These truncations cause additional uncertainties, as we have discussed in Refs.~\cite{BiLa13,HeBo13,RoBi12}.

In conclusion, a systematic quantification of the uncertainties inherent to the Hamiltonian remains one of the prime challenges of ab initio nuclear structure theory. Here we have started to address uncertainties related to the SRG-transformation and the various technical truncations of the Hamiltonian. Now that these uncertainties are understood, one can start to address the uncertainties related to the chiral EFT input itself. A systematic propagation of the uncertainties of the low-energy constants and uncertainties due to omissions of higher-order contributions in the chiral power counting will be the subject of future studies. It is evident already, that providing rigorous theoretical uncertainties for nuclear structure observables is at least as challenging as performing the ab initio calculation in the first place.

\section*{Acknowledgments}

We thank Petr Navr\'atil for many helpful discussions and for providing us with the \textsc{ManyEff} code. Supported by the Deutsche Forschungsgemeinschaft through contract SFB 634, by the Helmholtz International Center for FAIR (HIC for FAIR) within the LOEWE program of the State of Hesse, and the BMBF through contract 06DA7047I. 
Numerical calculations have been performed at the computing center of the TU Darmstadt (lichtenberg), at the J\"ulich Supercomputing Centre (juropa), at the LOEWE-CSC Frankfurt, and at the National Energy Research Scientific Computing Center supported by the Office of Science of the U.S. Department of Energy under Contract No.~DE-AC02-05CH11231. 


\end{document}